\documentclass[pre,
 reprint,
 amsmath,amssymb,
 aps,twocolumn, showkeys,
]{revtex4}
\usepackage{wasysym}
\usepackage{color}
\usepackage{graphicx}
\usepackage{dcolumn}
\usepackage{bm}
\usepackage{epsfig,subfigure,float,pseudocode,multirow,footmisc,rotating}
\usepackage[nottoc, notlof, notlot]{tocbibind}
\usepackage{hyperref}
\usepackage{rotating}

\begin{document}

\title{Loose social organisation of AB strain zebrafish groups in a two-patch environment}


\author{Axel S\'eguret}
 \email{axel.seguret@univ-paris-diderot.fr}
\affiliation{
 Univ. Paris Diderot, Sorbonne Paris Cit\'e\\
 LIED, UMR 8236, 75013, Paris, France
}

\author{Bertrand Collignon}
\affiliation{
 Univ. Paris Diderot, Sorbonne Paris Cit\'e\\
 LIED, UMR 8236, 75013, Paris, France
}

\author{L\'eo Cazenille}
\affiliation{
 Univ. Paris Diderot, Sorbonne Paris Cit\'e\\
 LIED, UMR 8236, 75013, Paris, France
}

\author{Yohann Chemtob}
\affiliation{
 Univ. Paris Diderot, Sorbonne Paris Cit\'e\\
 LIED, UMR 8236, 75013, Paris, France
}

\author{Jos\'e Halloy}
 \email{jose.halloy@univ-paris-diderot.fr}
\affiliation{
 Univ. Paris Diderot, Sorbonne Paris Cit\'e\\
 LIED, UMR 8236, 75013, Paris, France
}

 
\begin{abstract}
We explore the collective behaviours of 7 group sizes: 1, 2, 3, 5, 7, 10 and 20 AB zebrafish (\textit{Danio rerio}) in a constraint environment composed of two identical squared rooms connected by a corridor. This simple set-up is similar to a natural patchy environment. We track the positions and the identities of the fish and compute the metrics at the group and at the individual levels. First, we show that the size of the population affects the behaviour of each individual in a group, the cohesion of the groups, the preferential interactions and the transition dynamics between the two rooms. Second, during collective departures, we show that the rankings of exit correspond to the topological organisations of the fish prior to their collective departure with no leadership. This spatial organisation emerge in the group a few seconds before a collective departure. These results provide new evidences on the spatial organisation of the groups and the effect of the population size on individual and collective behaviours in a patchy environment.
\end{abstract}

\keywords{Collective behaviour --- Collective transition --- Zebrafish --- Collective departure}

\maketitle

\section{Introduction}

Across the collective behaviours observed in animals, collective movements \cite{Petit.2010, Sueur.2011a, Radakov.1973, Parrish.2002, Engeszer.2007, Herbert-read.2011, Hemelrijk.2012, Sueur.2011b}, nest site selections \cite{Franksetal.2002, Ame.2006, Rieucau.2010, Dahlbom.2011} and site transitions \cite{Harcourt.2009} have been evidenced in many species. In the latter case, the animals alone or in group face several alternatives and transit between them. The study of these transitions relies on decision-making processes and individual or collective preferences for environmental \cite{Conradt.2005} or group members characteristics \cite{Pritchard.2001, Engeszer.2004, Hoare.2004, Engeszer.2007} like leadership \cite{Couzin.2005}, motion \cite{Bourjade.2009} or behaviour, for example bold and shy individuals \cite{Dahlbom.2011, Leblond.2006}.

Numerous animal species have been observed in different sorts of constraint setups or mazes to study collective movement from one site to another: corridor type \cite{Pettersson.1993, Engeszer.2004, Engeszer.2007}, Y-maze \cite{Ward.2011}, T-maze \cite{Kistler.2011} or Plus-maze \cite{Sison.2010, Miller.2013}. Such constraint set-ups engage the animals to transit alone or in group from site to site and allow the observation of leadership \cite{Ward.2013, Bourjade.2015}, initiation of group movements \cite{Bourjade.2009, Ward.2013, Rosenthal.2015}, followers organisations \cite{Ward.2013}, pre-departure behaviours \cite{Bourjade.2009, Bourjade.2015} and sites transitions \cite{Harcourt.2009, Nakayama.2011, Nakayama.2012a}. In these latter cases the authors studied the transitions from one site to the other of one and two fish separated by a transparent partition (\textit{Gasterosteus aculeatus} and \textit{Sciaenops ocellatus}). Although such experimental procedure provided evidence of different leader/follower behaviour in fish, they prevent the fish from direct interactions between each other during the departures.

Studies performed with groups of fish swimming together have evidenced that the population sizes can impact swimming behaviours with variety of results. \cite{Becco.2006, Herbert-Read.2013, Tunstrom.2013} showed that the speed, the turning speed, the nearest neighbour distances, the milling or the alignment are affected by the number of group members. The authors present opposite results depending on the species: increasing the group size of \textit{Oreochromis niloticus} (330 and 905 fish), makes stronger alignments \cite{Becco.2006}, when for \textit{Notemigonus crysoleucas} (30, 70, 150 and 300 fish) alignments decrease \cite{Tunstrom.2013}.

Rather than investigating the influence of the group size on the swimming characteristic of the group, here, we focus on the collective movements between two environmental patches. In particular, we would like to characterise the dynamics of departure during sites transitions for several population sizes (1, 2, 3, 5, 7, 10 and 20 individuals) of AB zebrafish swimming in a constraint environment. Zebrafish are a gregarious vertebrate model organisms often used in behavioural studies \cite{Norton.2010, Oliveira.2013}.

In the laboratory as much as in the nature, the zebrafish behave in groups \cite{Engeszer.2007, Spence.2008, Mcclure.2006}. They come from India and live in small groups or in big shoals of several hundreds of fish depending on the region and the water or the environmental features (temperature, pH, human activity, predators, ...) \cite{Pritchard.2001, Parichy.2015, Suriyampola.2016}. Zebrafish live in a wide variability of habitats with varying structural complexities \cite{Arunachalam.2013, Suriyampola.2016} (from river channels, irrigation canals to beels) and we based our experimental method on the observations of fish swimming in a constraint set-up composed of two identical squared rooms connected by a long corridor evoking irrigation canals \cite{Spence.2006} and patchy environments \cite{Wiens.1976}.

We showed in a previous study that zebrafish transit from one landmark to another in an open environment during trials of one hour \cite{Seguret.2016}. Moreover, we showed that groups of fish were swimming along the border of the tank and thus had a strong thigmotactic tendency \cite{Collignon.2016}. Here, we introduce a new type of set-up to study group transitions following group departures. We vary the size of the population to see how the zebrafish will face this environment and adapt their individual and group behaviours. We also perform trials of one hour that allows to study a large number of repetitive transitions.


\section{Results}

\subsection{From the individual characteristics to pair interactions}

\begin{figure}[ht]
\centering
\includegraphics[width=0.5\textwidth]{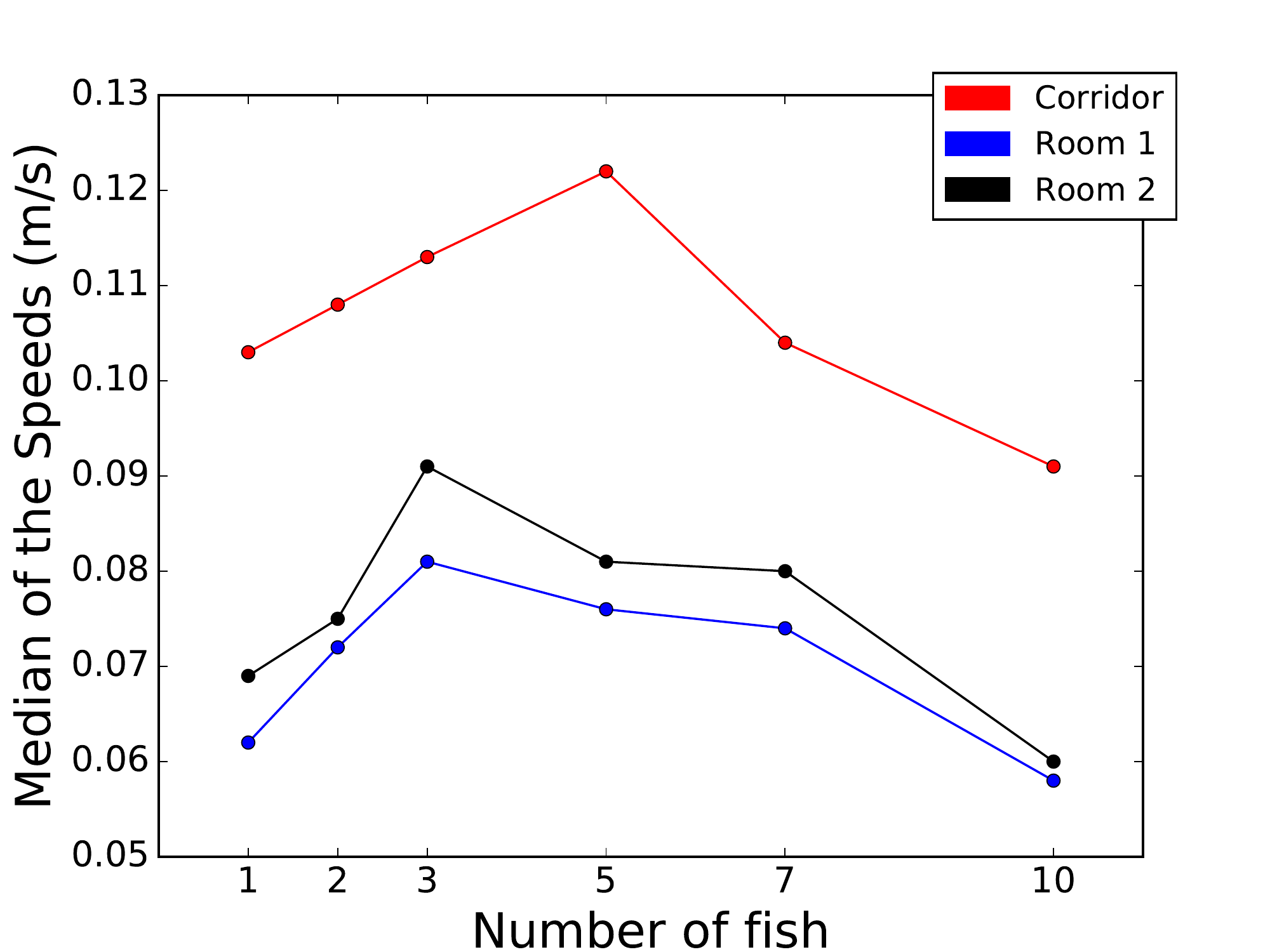}
\caption{\textbf{Medians of the individual speed distributions for different group sizes.} The red line represents the median of the individual speed for all individuals in the corridor, the blue line for the room 1 and the black line for the room 2. The zebrafish move faster in the corridor. In rooms 1 and 2, their speeds are similar. Groups of 3 zebrafish show the highest median speed in all areas.}
 \label{fig:median_speed}  
\end{figure}

The results presented in this paragraph focus on individual measures in 6 group sizes (1, 2, 3, 5, 7 and 10 fish). Figure~\ref{fig:median_speed} shows the medians of the individual speed of the fish measured during the entire experimental time (one hour) and according to their spatial location (in the corridor or in one of the two rooms). The fastest individuals are observed in groups of 5 fish in the corridor and 3 fish in both rooms. On the contrary, fish alone and groups of 10 individuals show the slowest median speeds. Moreover in the corridor, between the population sizes of 1 and 5 individuals, there is an increase of the speed median. Then, for larger population sizes, the medians decrease. Likewise, in both rooms, for the population sizes of 1 and 3, we observe an increase of the median of the speeds, then for bigger population sizes a drop.
First, we compared for each group size tested independently the speed of the individuals according to their location (in the corridor, in room 1 or in room 2). For all the group sizes, the speed of the individuals was significantly higher in the corridor than in the two rooms (for each group size : Kruskal-Wallis, $p < 0.001$, Tukey's honest significant difference criterion, $p < 0.001$ for corridor versus room 1 and $p < 0.001$ for corridor versus room 2). Indeed, the fish increase their swimming speed by approximately 3.5 cm/s in the corridor. They simply use the corridor to transit rapidly from room to the other. The difference between the speed measured in room 1 and room 2 was also significantly different for all the group sizes (Tukey's honest significant difference criterion, $p < 0.001$ for room 1 versus room 2). Although in this case, the difference between the swimming speed was always lower than 1 cm/s. Secondly, we compared the speed of the individuals in similar areas but for different group sizes. For each area (corridor, room 1 and room 2), the speed of the fish differs between all group sizes (for each area: Kruskal-Wallis, $p < 0.001$, Tukey's honest significant difference criterion, $p<0.001$ for all comparison) except in the corridor for groups of 1 and 7 fish. These first results show that both the group size and the location of the fish have an influence on their individual speed.

\begin{figure}[ht]
\centering
\includegraphics[width=0.50\textwidth]{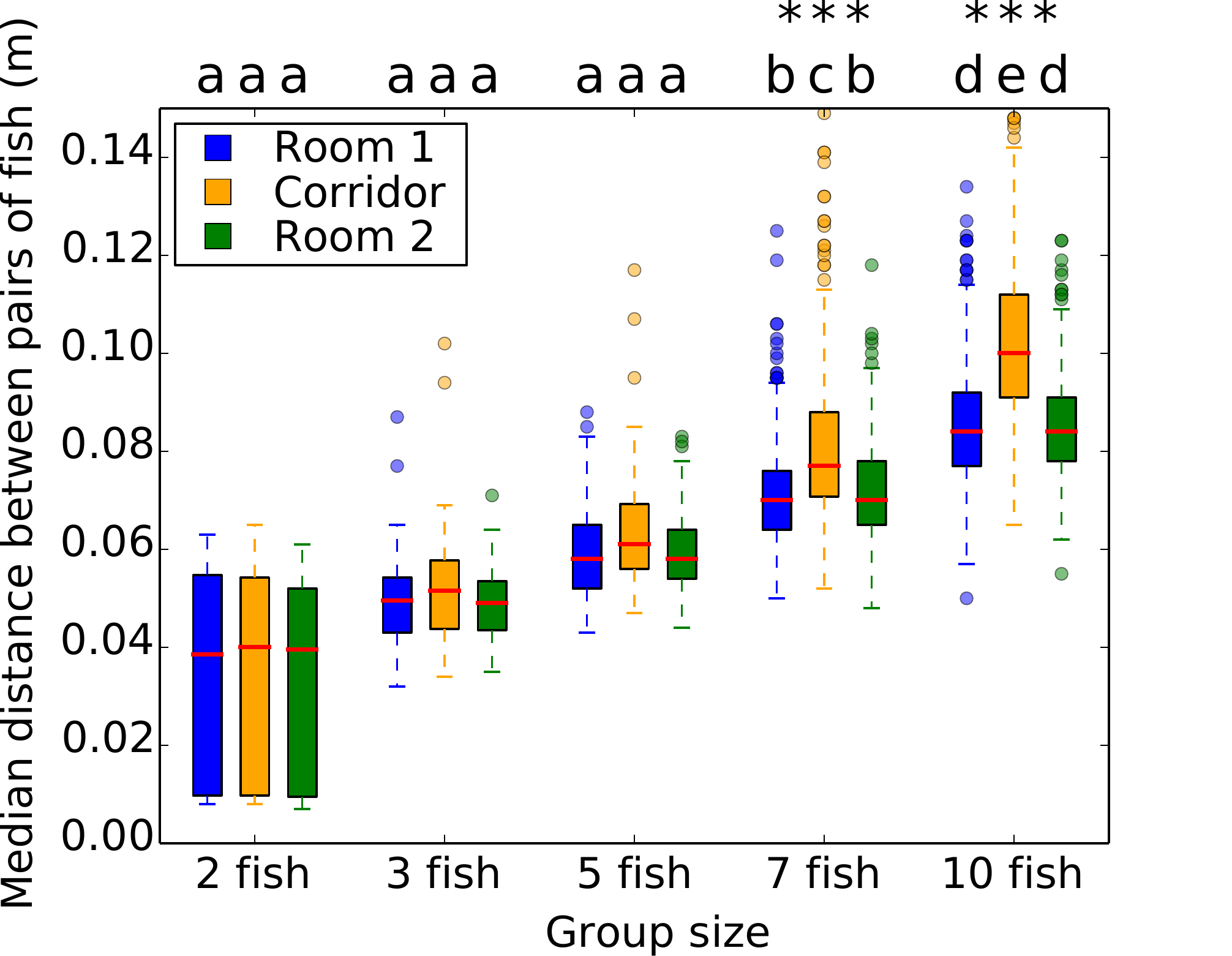}
 \caption{\textbf{Boxplots of the medians of the distance distribution between all respective pairs of zebrafish} (A) in the room 1, (B) in the room 2 and (C) in the corridor. The red line is the median. The bigger the population, the larger the distances between fish pairs. * = $p < 0.05$, ** = $p < 0.01$, *** = $p < 0.001$, ns = non significant.}
 \label{fig:couples_boxplot}
\end{figure}

Then, we study the evolution of the group structure according to the location and group size by measuring the inter-individual distances between each pair of individuals. As the tracking software is able to individually recognise the different group members during the entire experiment, we were able to measure the distance between each specific pair of fish from the beginning to the end of the observation period. These distances are presented in figures S\ref{fig:median_distances_pairs_1}, S\ref{fig:median_distances_pairs_2} and S\ref{fig:median_distances_pairs_corridor} of the appendix that visually represent the median distances between the group members (all pairs of fish) in room 1, room 2 and the corridor for all group sizes (2, 3, 5, 7 and 10 zebrafish) and trials (12). Figure~\ref{fig:couples_boxplot} shows the boxplots of these median distances between pairs of fish. Thus, the boxplots for 2 fish consists of 12 medians (12 x 1 couple), for 3 fish 36 medians (12 x 3 couples), for 5 fish 120 medians (12 x 10 couples), for 7 fish 252 medians (12 x 21 couples) and for 10 fish 540 medians (12 x 45 couples).
\\Again, we compared the distributions of the median distances between the pairs focusing on each area (room 1, room 2 and corridor) or each population size. In each area, the median distances do not significantly differ between groups of 2, 3 or 5 fish. On the contrary, fish group size of 7 show significantly larger distance from each other as well as fish in groups of 10 that differ from all other group sizes (Kruskall-Wallis $p < 0.001$, Tukey's honest significant difference criterion, $p > 0.05$ between groups of 2, 3 and 5 fish, $p < 0.001$ in all pair comparisons with groups of 7 and 10). Then, we compared for each group size the distribution of the distances according to the location of the fish. In this case, the distributions between both rooms are not significantly different for all group sizes but they are significantly different between the corridor and the rooms 1 or 2 for population sizes of 5, 7 and 10 individuals. The structure of the group is thus influenced by both the group size and the location of the fish in their environment.

\subsection{Oscillations and collective departures}

\begin{figure}[ht]
\centering
\includegraphics[width=0.45\textwidth]{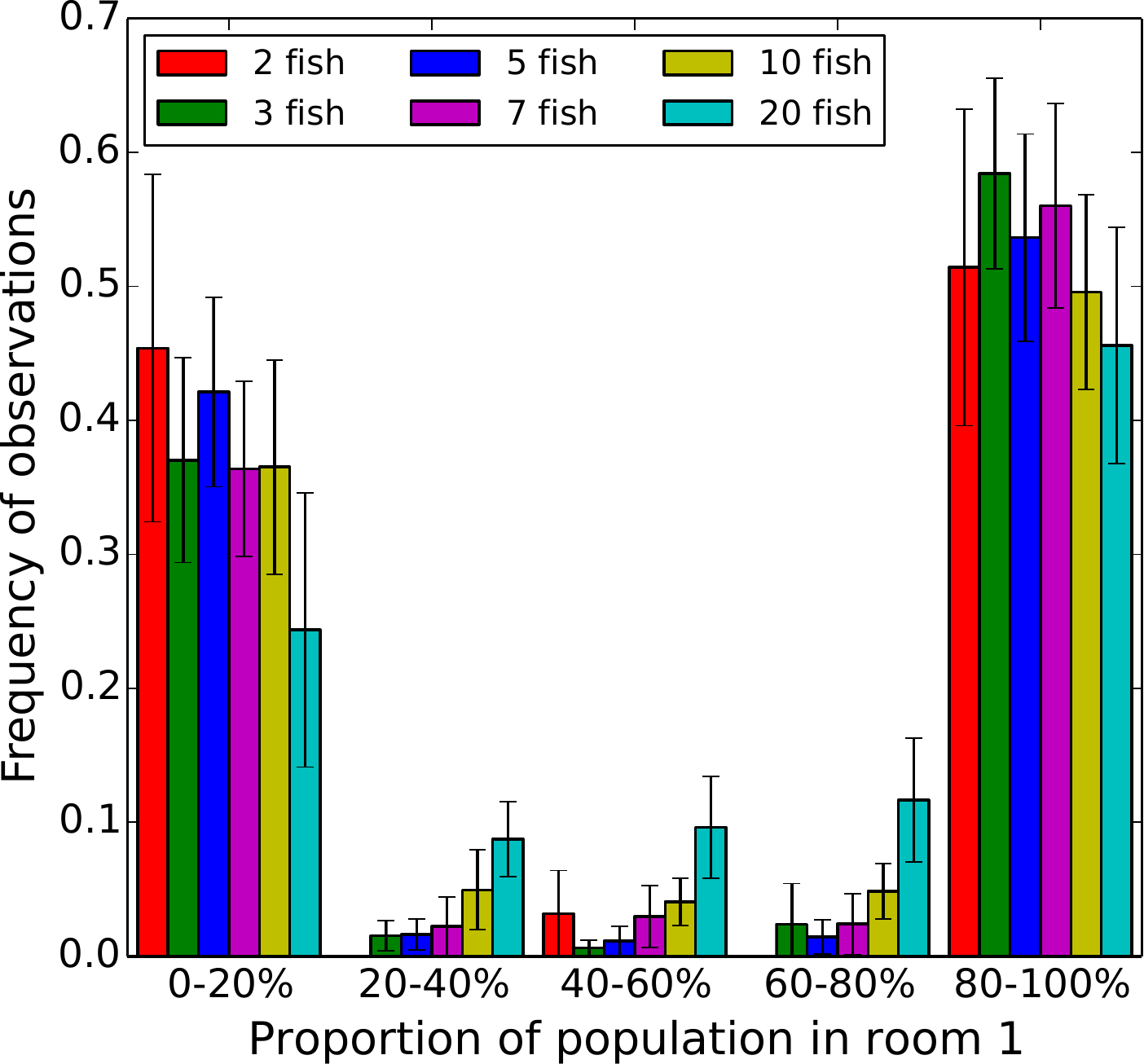}
 \caption{\textbf{Frequency of the proportion of the population in room 1.} Almost 35\% of the time, 0 to 20\% of the population is present in the room 1 when almost 50\% of the times, 80 to 100\% of the population is in the room 1. Focusing on more equal repartition of the fish between the rooms (40 to 60\% of the population), larger populations lead to higher frequency of group splitting.}
 \label{fig:frequence_presence_room1}  
\end{figure}

In this section, we characterise the collective behaviour of the fish. In particular, we focus our investigation on the oscillations between both rooms and the collective departure dynamics of the groups. First, we studied the repartition of the fish among the two rooms. Approximately 70\% of the positions of the fish were detected in the rooms, independently from the group size (Figure~\ref{fig:presence_rooms}). At each time step with at least one fish detected in a room, we analysed the repartition of the group among the rooms by computing the proportion of fish present in room 1 = R1 / (R1+R2) with $R_i$ the number of fish in room \textit{i}. In the Figure~\ref{fig:frequence_presence_room1}, we show that 80\% of the time, less than 20\% or more than 80\% of the population is detected in the room 1. This result highlights that, as expected for a social species, the fish are not spread homogeneously in the two rooms but aggregate collectively in the patches, with only few observations of homogeneous repartition in both rooms. However, this analysis also show that the proportion of observation with equal repartition between both rooms (40-60\%) increases with the size of the group. Thus, even if they are mainly observed together, fish in large group have a higher tendency to segregate in two groups. We show that the frequencies of observations for the proportions of 80 to 100\% of the population in room 1 are higher than 50\% for all size of the population, except for 10 and 20 fish. For each trial, we defined the room 1 as the starting room where we let the fish acclimatize during 5 minutes in a transparent perspex cylinder. This may explain the observed bias of presence in favour of room 1 that may be a consequence of longer residence time at the beginning of the trials.

\begin{figure}[ht]
\centering
\includegraphics[width=0.55\textwidth]{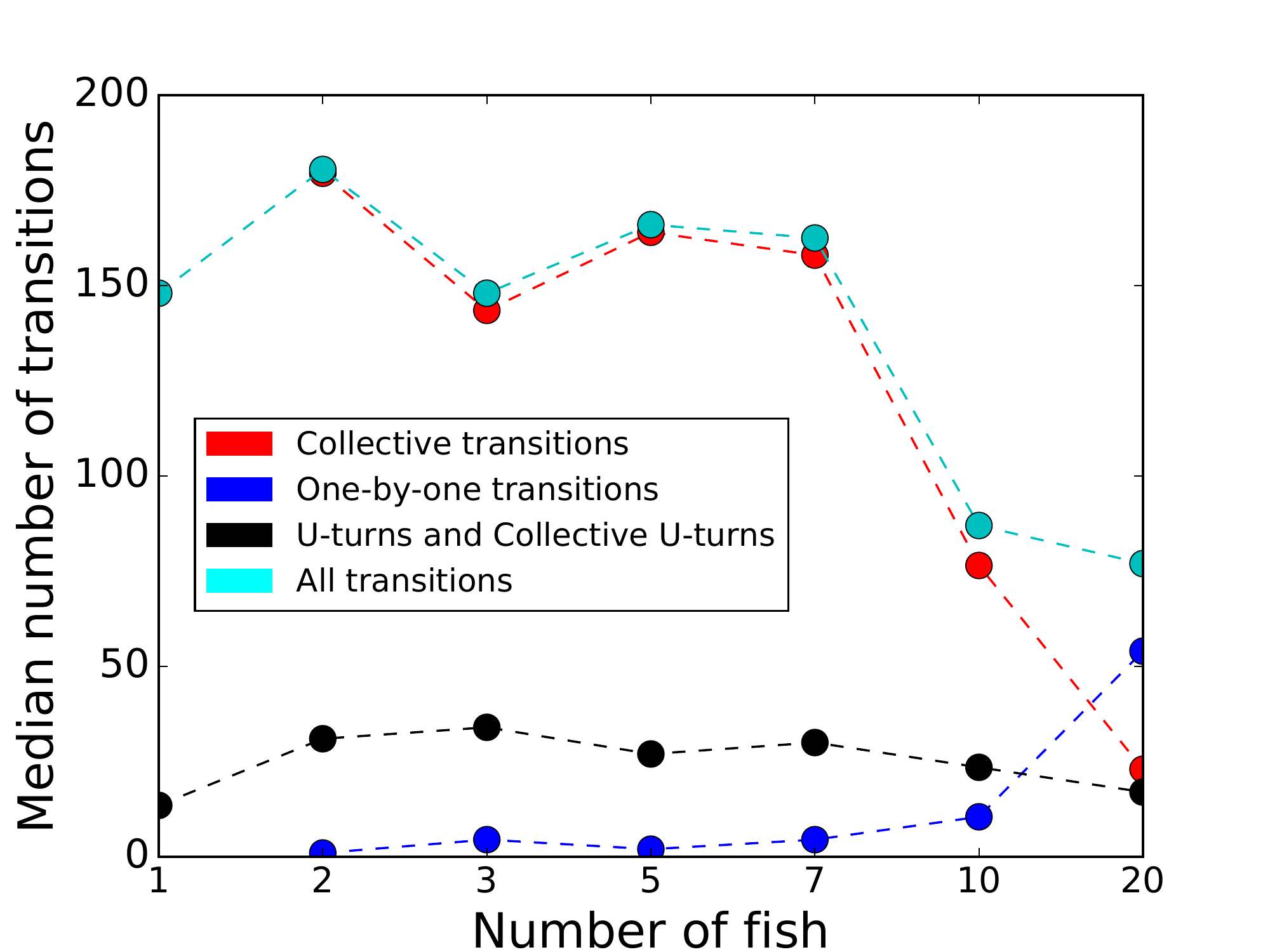}
 \caption{\textbf{Median number of transitions for different group sizes.} The red curve shows "Collective transitions", the blue curve shows "One-by-one transitions", the black curve represents the "Collective U-turns" and the magenta ("All transitions") is the sum of "Collective transitions" and "One-by-one transitions". "One-by-one transitions" occur when the fish transit one by one from one room to the other. "Collective transitions" appear when the group transit between both rooms through the corridor. "Collective U-turns" occur when the group go back to the previous room. The dashed lines facilitate the lecture. The figure shows that increasing the group sizes makes the number of "Collective U-turns" and "Collective transitions" decrease and the number of "One-by-one transitions" increase. Each point shows the median of 12 values.}
 \label{fig:transitions}  
\end{figure}

Then, since the fish are observed most of the time forming one group in one of the two rooms, we studied the transitions of the majority of fish between the two patches during the whole experiment. In the Figure~\ref{fig:transitions}, we plot the median numbers of transitions between both rooms (see Figure~S\ref{fig:mean_transitions} and Table~S\ref{fig:std_events} (D) of the appendix for the plot of the means of the numbers of transitions and their standard deviations in a table). We identify three categories of transitions. "One-by-one transitions" occur when the fish transit one by one from one room to the other, "Collective transitions" appear when the group transit between both rooms through the corridor and "Collective U-turns" occur when the group go back to the previous room. For 1 zebrafish, the "One-by-one transitions" and "Collective transitions" do not lake sense, thus we created another category called "All transitions". This last category is the sum of the "One-by-one transitions" and the "Collective transitions". Larger group sizes makes the number of "Collective U-turns", "Collective transitions" and "All transitions" decrease and the number of "One-by-one transitions" increase. For the transitions (collective, one-by-one and all), this tendency intensifies for bigger groups of 10 and 20 zebrafish. Also, for groups of 3 zebrafish, there are less "Collective transitions" (as well as "All transitions") than for groups of 2, 5 and 7 zebrafish. "U-turns" and "Collective U-turns" stay rare and are very stable for all group sizes and their highest mean numbers are reached for group of 2 and 3 zebrafish. "One-by-one transitions" are as well very rare for small groups and increase when the group size reaches 10 zebrafish.
\\For each group size, we compared with a Kruskal-Wallis test the distributions of the number of transitions (Collective, One-by-one and U-turns) and found a $p < 0.001$, which shows that at least one of the distributions is significantly different from the others. The Tukey's honest significant difference criterion shows that: all the distributions are significantly different except in groups of 10 individuals between "Collective U-turns" and "One-by-one transitions" and in groups of 20 individuals between "Collective U-turns" and "Collective transitions".

\begin{figure}[ht]
\centering
\includegraphics[width=0.5\textwidth]{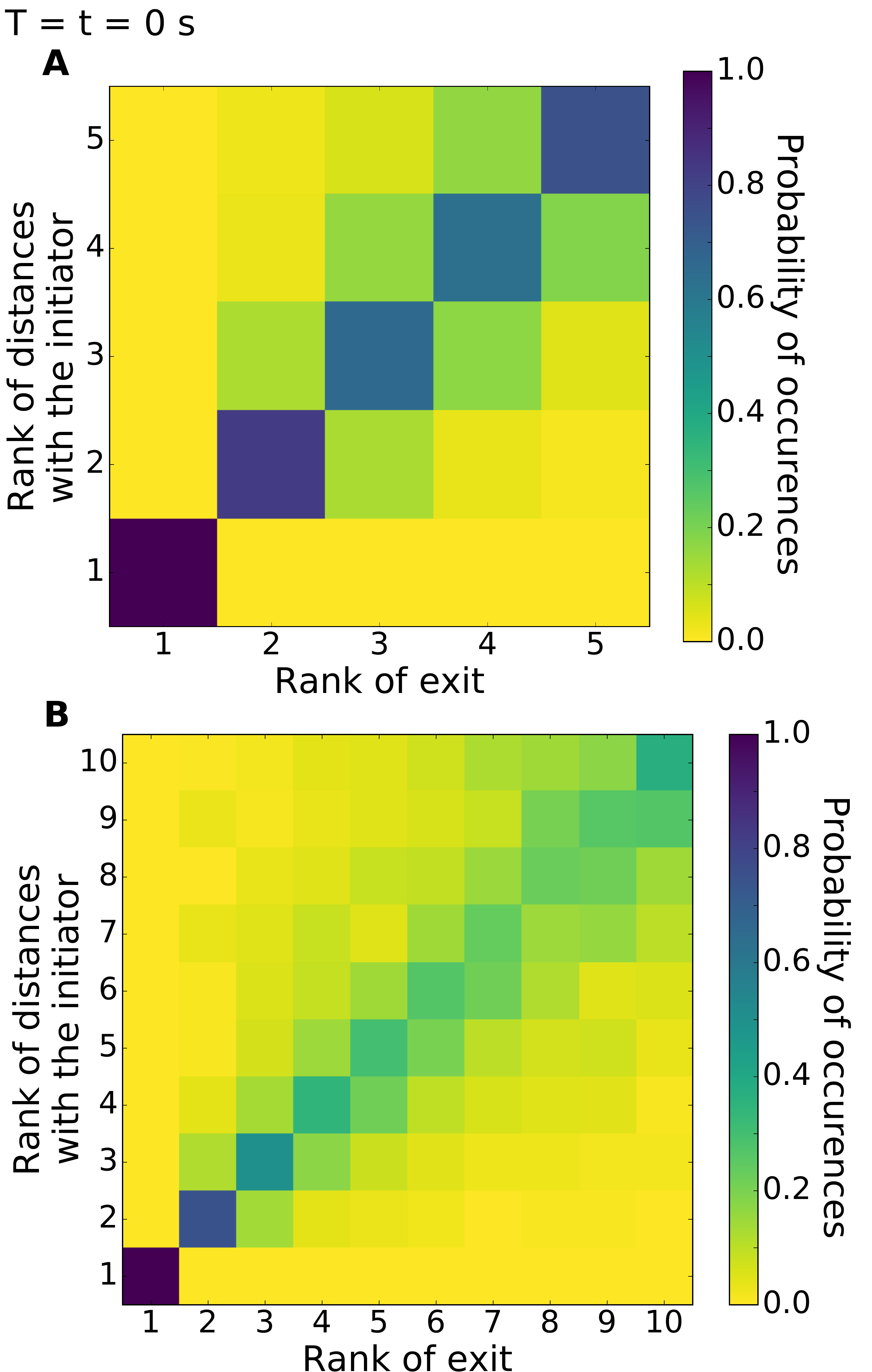}
 \caption{\textbf{Probability of occurrence of the rank of exit with the rank of distance to the initiator} for population sizes of 5 zebrafish (left column) and 10 zebrafish (right column). We counted N = 1456 exits for 12 replicates with 5 zebrafish and N = 277 for 12 replicates with 10 zebrafish. (A) and (B) show the map at the time when the initiator leave the room. As an example, in (A) the probability of occurrence where the second fish leaves the room and has the shortest distance from the initiator is 0.82. This probability decreases to 0.12 for fish with rank of 2 for exit and rank of 3 for distances (the second closest distance from the initiator).}
 \label{fig:map_sortie}
\end{figure}

\begin{figure*}[ht]
\centering
\includegraphics[width=0.7\textwidth]{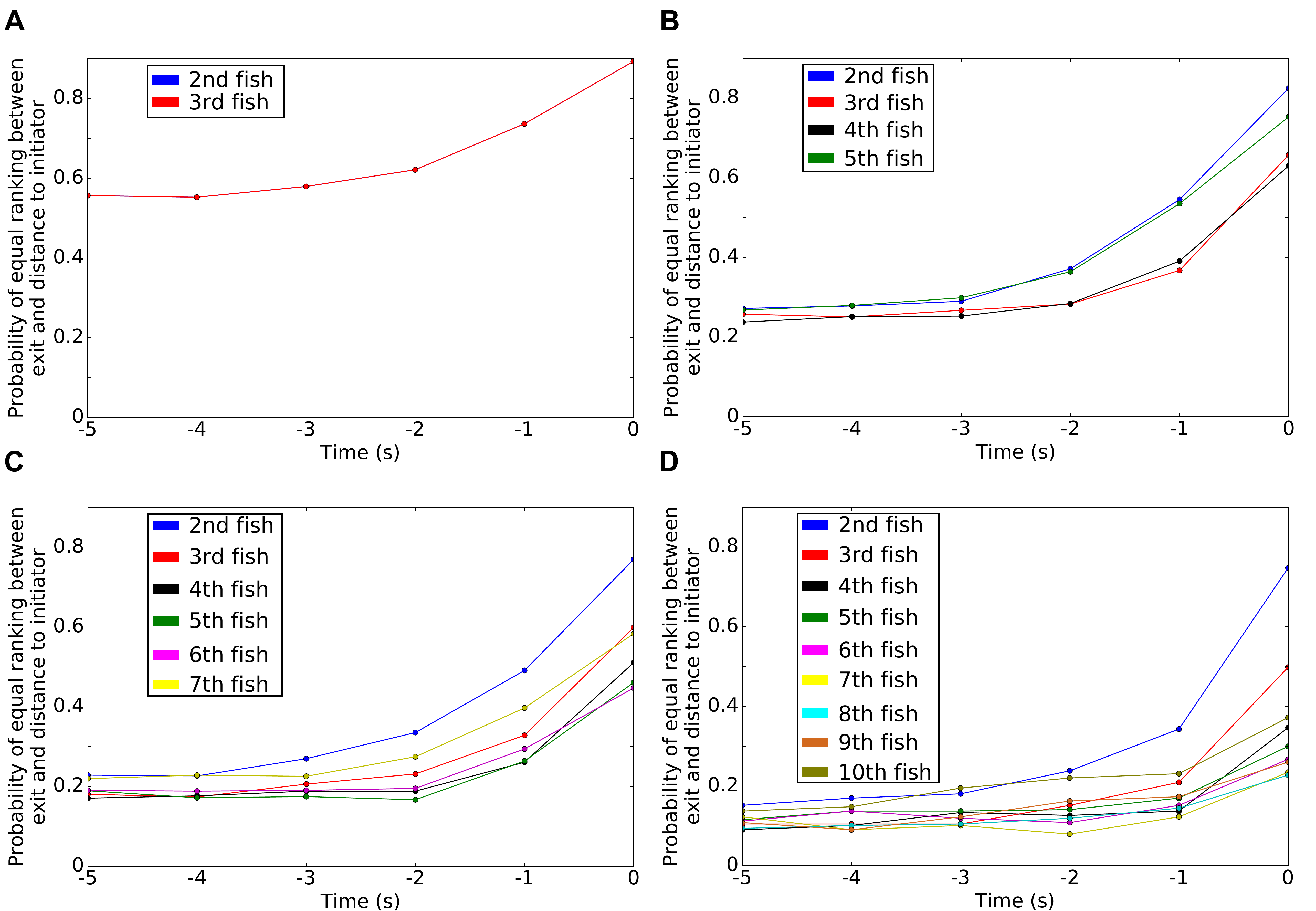}
 \caption{\textbf{Time series of the probability of equal ranking between the rank of exit and the rank of distances from the initiator} for groups of (A) 3 zebrafish, (B) 5 zebrafish, (C) 7 zebrafish and (D) 10 zebrafish. This figure is related to the results shown in Figure~\ref{fig:map_sortie} (the diagonal). We plot a time series of the 5 seconds before the initiation. We show that the probability increases strongly 2 seconds before the initiation. We see also that this probability is the highest for the first ranked fish and higher for the 3 first ranked fish and for the last ranked fish. This behaviour is also valid even few seconds before the initiation.}
 \label{fig:plot_sortie}
\end{figure*}

\begin{figure*}[ht]
\centering
\includegraphics[width=.45\textwidth]{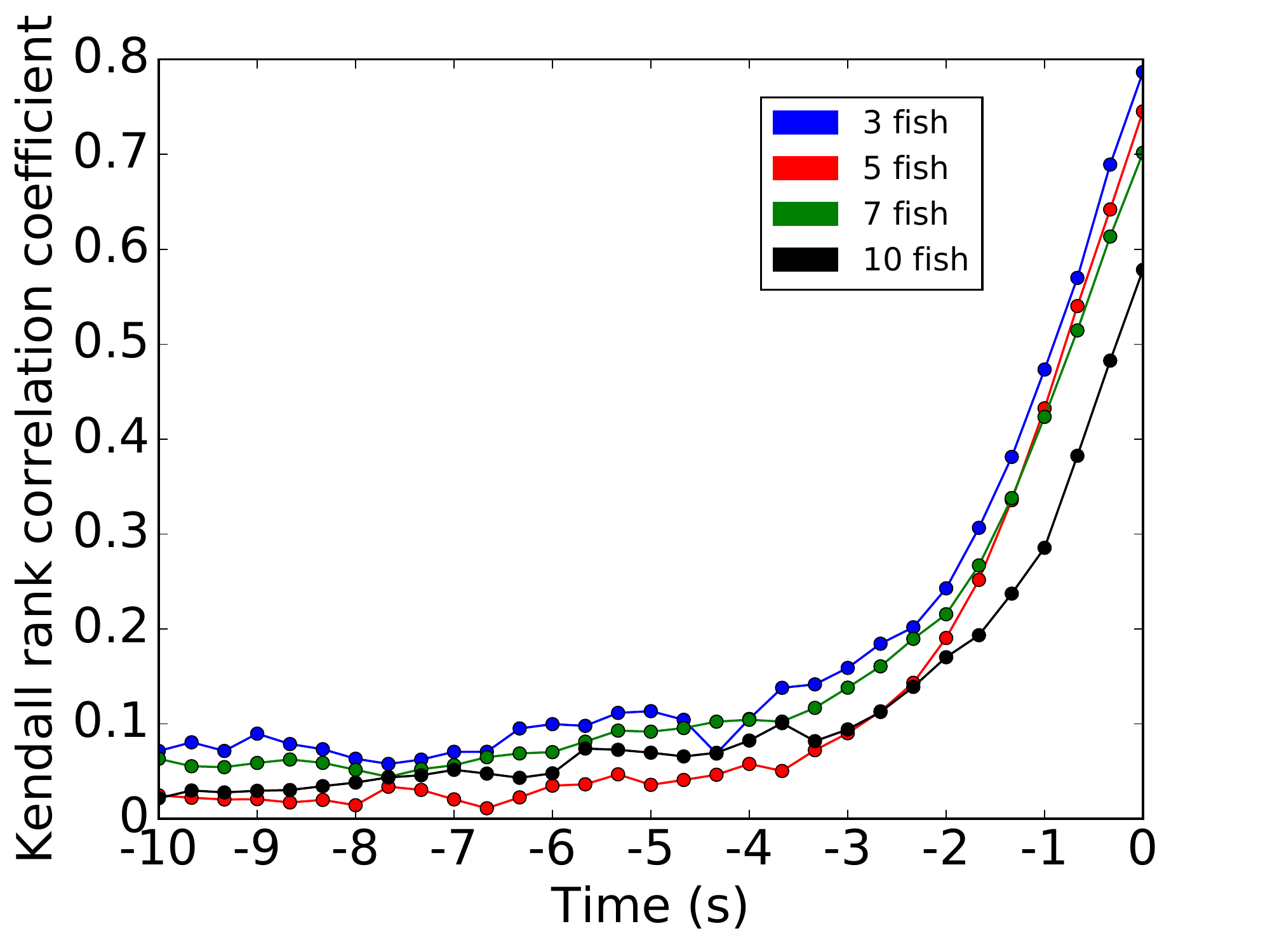}
 \caption{\textbf{Time series of the Kendall rank correlation coefficient} calculated on the results of the Figures~S\ref{fig:map_sortie_5_10} and ~S\ref{fig:map_sortie_3_7}. It has been calculated every 1/3 second starting 10 seconds before the initiation. The Kendall rank correlation coefficient is a measure of ordinal association between two measured quantities. It goes to 0 when the two quantities are independent and goes to 1 if they are correlated. For example with 5 zebrafish, the time series shows that from 4 seconds before the initiation the Kendall rank increases from 0.06 to 0.75 (at T = t = 0 s). Whatever the size of the groups, we conclude that the closer to the initiation the higher the correlation between the rank of exit and the rank of the distances with the initiator.}
 \label{fig:kendall_rank}
\end{figure*}

As most of the transitions occur in groups, we analysed the dynamics of collective departure from the rooms. Thus, for each collective departure of the fish, defined as the whole group leaving one of the resting sites for the corridor towards the other one, we identified the ranking of exit of each fish and also their distance from the first fish leaving the room (i.e. defined as the initiator). Figure~\ref{fig:map_sortie} represents the normalised contingency table of the rank of exit for all zebrafish from both rooms (without distinction) with the rank of the distances of all zebrafish to the initator. These results correspond to 12 replicates of groups of 5 and 10 zebrafish. The initiator has a rank of exit and a rank of distances of 1.  For example, in (A) the probability that the first fish to follow the initiator (rank 2) was also the closest fish of the initiator when it exited the room is 0.82. Figure~S\ref{fig:map_sortie_5_10} and ~S\ref{fig:map_sortie_3_7} of the appendix show a more detailed version of the Figure~\ref{fig:map_sortie} for 3, 5, 7 and 10 individuals. In Figure~\ref{fig:plot_sortie}, we plot for 3, 5, 7 and 10 zebrafish the values of the probability of equal ranking between the exit and the distances with the initiator (i.e. the diagonal of the previous plots -- Figure~\ref{fig:map_sortie}) for different time-lag before the exit of the initiator. In particular, we computed the ranking of the distance from the initiator at 1 to 5 seconds before the exit of the initiator. First, these measures show that the further from the time of the initiation the lower the probability of equal ranking. This assessment is valid for every group sizes. Second, we see that the probability of equal ranking is often higher for the first and for the last ranked fish even few seconds before the initiation (around 2 seconds before the initiation). In Figure~\ref{fig:kendall_rank}, we use the Kendall rank correlation coefficient to see if the rank of exit and the rank of distances with the initiator are dependant (close to 1) or not (close to 0) through the time. For every group sizes, we show an increase of the Kendall rank correlation coefficient when closer to the initiation. For 3 zebrafish, the time series shows that from 4 seconds before the initiation the Kendall rank correlation coefficient fully increases from 0.11 to 0.79 (at T = t = 0 s). For 5 zebrafish, it increases from 0.06 (at T = t - 4 s) to 0.75 (at T = t = 0 S), for 7 zebrafish, it increases from 0.10 to 0.70 and for 10 zebrafish, it increases from 0.08 to 0.58. These results show that for all group sizes, the closer to the initiation the higher the correlation between the rank of exit and the rank of the distances with the initiator.

\clearpage


\section{Discussion}

We studied the impact of the group size (1, 2, 3, 5, 7, 10 or 20 individuals) on the collective motion and the collective departure between two environmental patches in adult AB zebrafish. We show that the individual speed of the zebrafish varies according to the areas in which they are swimming and their population size. The zebrafish move faster in the corridor and have similar lower speeds in both rooms. The surface of the corridor is the third of a room and it constraints the direction the fish have to follow. We have shown that zebrafish are known to swim along the walls of the experimental tank \cite{Collignon.2016, Seguret.2016} thus showing a strong thigmotaxis.In the corridor, that can be compared to a tunnel, canalised by the walls of the corridor, the zebrafish increase their individual speeds to make the transit from one room to the other. In both rooms the medians of the individual speeds are at their highest levels for groups of 3 zebrafish and the maximum of the median of the individual speeds is reached for population size of 5 fish in the corridor. In parallel, in each area these medians are at their lowest levels for the smallest and the biggest population sizes: 1 and 10 zebrafish. Hence, in both rooms and in the corridor respectively, we have seen that from 1 to 3 individuals and from 1 to 5 individuals the individual speeds increase, when from 3 to 10 individuals and from 5 to 10 individuals, the individual speeds decrease. We showed also that the distances travelled by the zebrafish are related to the size of the population (Figure~S\ref{fig:cumul_distance} and Table~S\ref{fig:median_distance_compare} of the appendix). Groups of 2 to 7 zebrafish travelled the longer distances (with a declining trend) and fish alone and groups of 10 zebrafish travelled the shorter distances.

First, our results confirm that the behaviour of a zebrafish alone differs significantly from the behavior of zebrafish in groups. Isolated zebrafish travel a shorter distance and at a lower speed than zebrafish in groups. This can be the result of the stress generated by being isolated in a new environment. The stress level has been studied and \cite{Egan.2009} shows that some anxiolytics (fluoxetine and ethanol, that reduce stress level) will increase the speed and/or the travelled distances of zebrafish alone in a tank. Second, zebrafish swim faster in smaller population sizes and their speeds decrease for bigger population sizes. We observe the same trend for the travelled distances. These results may suggest a congestion effect where obstruction can affect their individual speeds and hence their travelled distances during the experimentation time \cite{Chowdhury.2005}. Such effect has already been reported for example in the ant species \textit{Atta cephalotes}: crowded conditions on the trail network make the velocity decrease \cite{Burd.2003}. Herbert-Read et al. present another explanation for the changes of the motions where each fish (\textit{Gambusia holbrooki}) conforms to the group behaviour through the interaction rules between the individuals and the decisions of each individual to follow or copy their neighbour movements \cite{Herbert-Read.2013}. Although this case seems to be extreme, \cite{Magnhagen.2009, Burns.2012, Nakayama.2012b} have shown that fish from different species (\textit{Perca fluviatilis}, \textit{Gasterosteus aculeatus}) and \textit{Gambusia holbrooki} can maintain particular individual behavioural traits in a social context. These changes in behaviours are found in other animal species such as birds (\textit{Erythrura gouldiae}) that adjust their behaviour according to the personality of their partners \cite{King.2015}.\\

\begin{figure}[ht]
\centering
\includegraphics[width=0.80\textwidth]{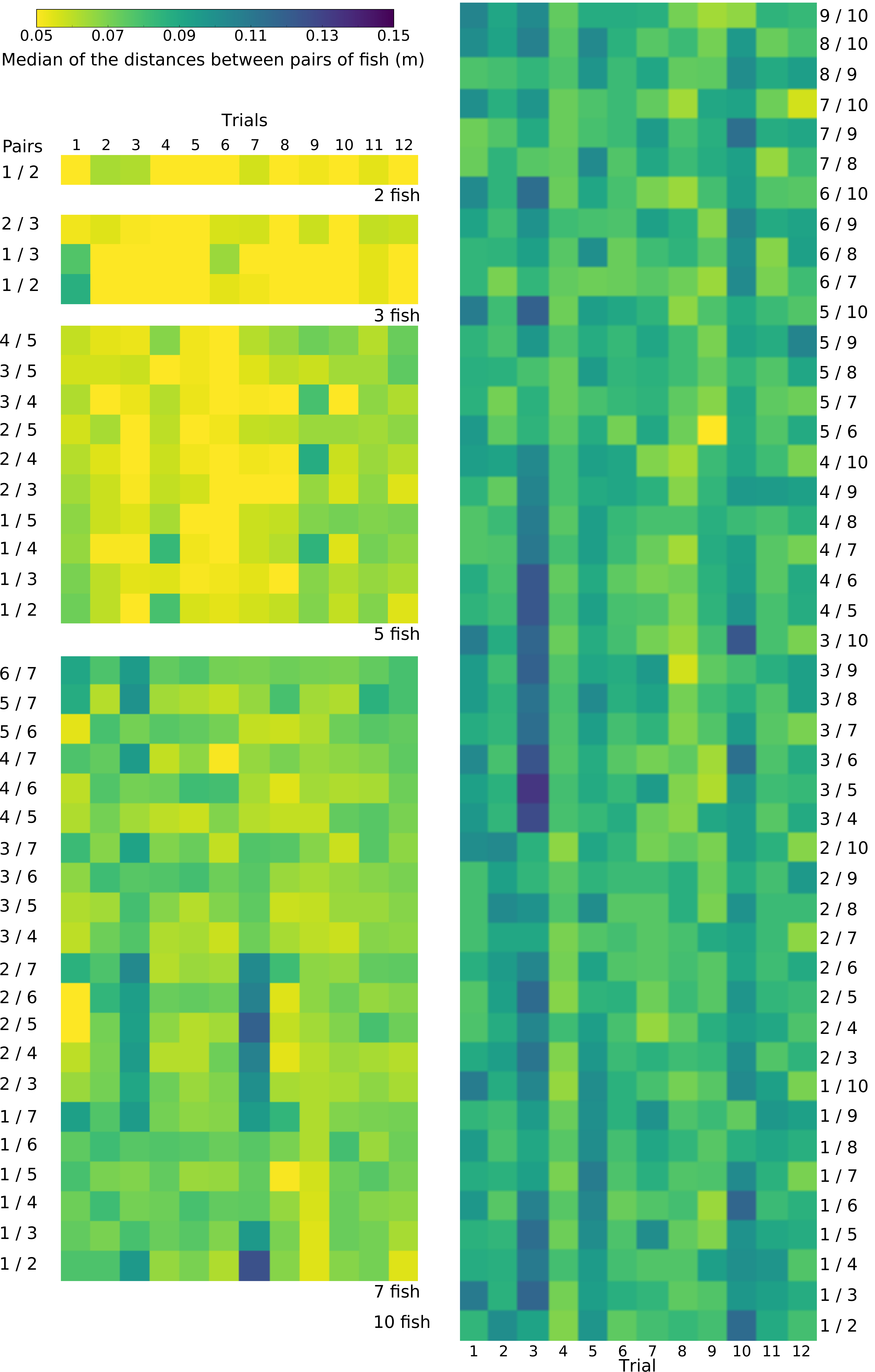}
 \caption{\textbf{Medians of the distances between all respective pairs of zebrafish per trial in the room 1 for 5 different group sizes.}  Each square corresponds to the median of the distances between the fish of one specific pair during a whole trial (out of 12 trials). When the colour of the square is yellow it means that the median of the distances between the fish is small (0.05m), when the colour is dark-blue it means that the median of the distances is larger (0.15m). For each area, we see a large distribution of the medians of the distances and their increase with the population size.}
 \label{fig:median_distances_pairs_1}
\end{figure}

By analysing the distances between all pairs within a group we show that the larger the population the higher the medians of the distances between all respective pairs of zebrafish (Figures~\ref{fig:couples_boxplot} and ~S\ref{fig:median_distances_pairs_1}, ~S\ref{fig:median_distances_pairs_2} and ~S\ref{fig:median_distances_pairs_corridor} of the appendix). In parallel (Figure~S\ref{fig:all_neighbour} of the appendix), if we focus on the neighbour distances, by increasing the size of the population (2 to 5 individuals) the median of the average neighbour distances decreases from 0.08m to 0.04m and stabilize around this value for groups of 7, 10 and 20 zebrafish. The combination of these results shows that there is a clear effect of the size of the population on the cohesion of the group; an effect that we already have shown in \cite{Seguret.2016} where the bigger the population the higher the cohesion of the whole group.\\

It seems that there are preferential interactions between zebrafish (Figure~S\ref{fig:median_distances_pairs_1}, ~S\ref{fig:median_distances_pairs_2} and ~S\ref{fig:median_distances_pairs_corridor} of the appendix) and increasing the size of the population will affect these interactions (Figure~\ref{fig:couples_boxplot}): respective pairs are less cohesive in marger groups. Preferential interactions have been evidenced in other species: Briard et al. \cite{Briard.2015} show affinities, hierarchy and pairs interactions in a group of domestic horses,  Sueur et al. and King et al. \cite{Sueur.2008, Sueur.2009, King.2011} show that the affinity between individuals (\textit{Macaca mulatta}, \textit{Macaca tonkeana} and \textit{Papio ursinus}) play a role in the collective movements. We propose two hypotheses that could explain the evolution of the interactions between pairs of zebrafish when changing the size of the population. On the one hand, in groups larger than 2, each zebrafish has to choose the preferred partners, between all other fish. Larger populations lead to make more individual choices and more preference tests. On the other hand, the patchy environment may break pair interactions and may force the emergence of new pairs. These two hypotheses could explain the dynamics of the pair interactions observed during the experiments.\\

\begin{figure}[ht]
\centering
\includegraphics[width=0.75\textwidth]{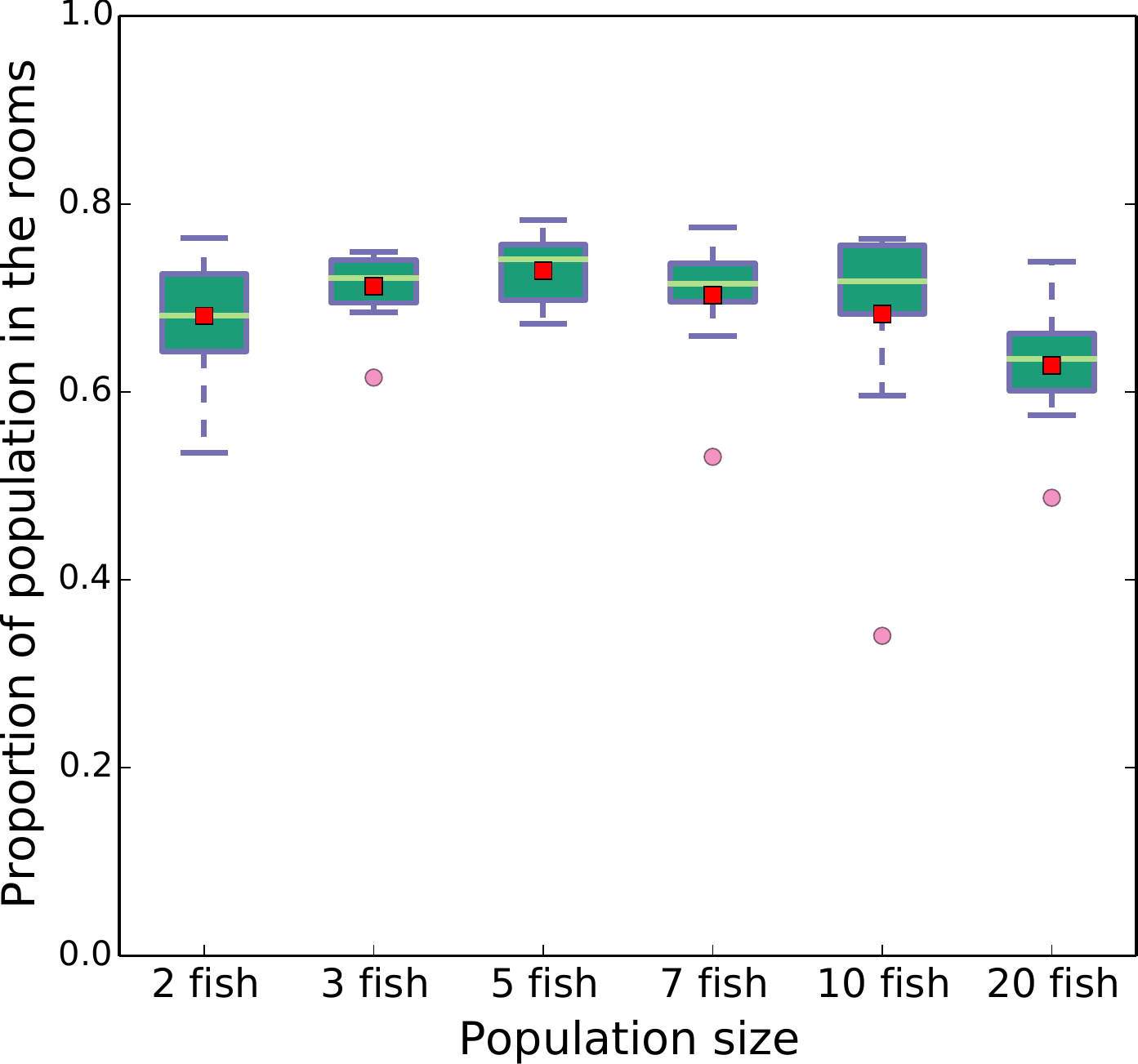}
 \caption{\textbf{Proportion of fish detected in both rooms.} We show that on average 70\% of the fish are detected in the rooms whatever the size of the population. The red square shows the mean and the lighter line the median.  We compared the distributions with Kolmogorov-Smirnov tests and found a $p-value < 0.001$ when comparing the distributions of population sizes of 2 fish versus 20 fish and 3 versus 20. The others comparisons of the distributions where always non significantly different.}
 \label{fig:presence_rooms}  
\end{figure}

The fish are detected 70\% of the time in the rooms (Figure~S\ref{fig:presence_rooms}). On average, they spend about 10 seconds in a room (Figure~S\ref{fig:durations}) then transit the the other one through the corridor (about 4 seconds on average) and then come back. They oscillate between the rooms. In a previous study we showed that zebrafish also transit and oscillate between landmarks in an open environment \cite{Seguret.2016}. Figure~\ref{fig:transitions} shows that most of the transitions are collective. Compared with Figure~\ref{fig:frequence_presence_room1} it shows that the whole population swim together in both rooms. This observation is strengthened by the very rare number of "One-by-one" transitions between the rooms. However, groups of 10 and 20 zebrafish show sharp decreases in the number of collective transitions. This drop could be due to the topology of the set-up and congestion effects. Larger groups can split into smaller sub-groups. The threshold we imposed in the analysis of the collective transitions (below 70\% of the whole population the transitions were not taken into account) may reinforce this effetc. This seems to be confirmed by the Figure~\ref{fig:frequence_presence_room1} that shows a larger occurence of fish distribution between both rooms for larger group size. Many studies have analysed the fusion-fission mechanisms occurring in groups of fish or mammalians. \cite{Hoare.2000, Krause.2002, Croft.2003} show that these mechanisms are frequent in the wild and generate body length assortment within groups of fish (\textit{Fundulus diaphanus}, \textit{Notemigonus crysoleucas}, \textit{Catostomus commersonii}, \textit{Poecilia reticulata}). Sueur et al. \cite{Sueur.2011b} show that fission-fusion mechanisms participate in the information transfer between sub-groups and the group of \textit{Myotis bechsteinii}.\\
In the corridor, we observe few u-turns. The zebrafish swimming preferentially along the walls and a canalisation effect of the corridor may explain this observation. The zebrafish also show higher speeds in the corridor. As expected, the corridor connecting the two patches is used as a mere transit area.\\

We show that the organisation of the group emerges during collective departure and is related to the distances between the initiator of the exit from the room and the other fish.  Ward \textit{et al.} has shown that the first fish (of a population of 5 \textit{Dascyllus aruanus}) to follow the initiator is generally (rank = 2: 53\% of the trials over 2 trials for each 15 groups of fish) the nearest neighbour of the initiator and that the frequency of equality between the rank of exit and the rank of the distances from the initiator decreases, with these results rank = 3: 27\% and 33\%, rank = 4: 20\% and 7\% then rank = 5: 0\% and 7\%) \cite{Ward.2013}. We tested 4 populations sizes (3, 5, 7 and 10 zebrafish) and show similar results especially on the decreasing trend of the probability when focusing on the next ranked fish Figures~S\ref{fig:map_sortie_3_7}, ~S\ref{fig:map_sortie_5_10} of the appendix and Figure~\ref{fig:plot_sortie}. However, we observe an extremely high probability of equal ranking for the first fish that follows the initiator (rank = 2: 75\% to 90\%), high probabilities of equal ranking for the second and the last fish that follow the initiator (rank = 3: 50\% to 65\% and rank = last fish: 38\% to 75\%) and show that the probabilities of equal ranking for the other fish are quite similar to each others. The rank of exit and the rank of the distances from the initiator are strongly correlated at the moment of the exit (T = 0s, Figure~\ref{fig:kendall_rank}), from 60\% to 80\%. Hence, it seems that the organisation of the zebrafish groups ($2^{nd}$, $3^{rd}$ and last ranked fish) during the collective departures is topological. Other studies about the organisations of collective departures show a joining process for \textit{Equus ferus caballus} that is related to affinities and hierarchical rank \cite{Briard.2015}. Rosenthal \textit{et al.} show that, in groups of \textit{Notemigonus crysoleucas}, the initiator is the closest fish from the group boundary in 27\% of the cases and the first responder is the closest fish from the group boundary in 19\% of the cases \cite{Rosenthal.2015}. Moreover during the initiation, when fish leave the rooms, our results suggest the idea of cascades of behavioural changes already developed by Rosenthal \textit{et al.} \cite{Rosenthal.2015}: the initiator drags another fish along that drags another one, etc.\\

This organisation appears  a few seconds before the fish leave a room to transit to the other one. Two seconds before the initiation, the group show a structure that prepares for the exit (Figure~\ref{fig:plot_sortie}). The Kendall rank correlation coefficient confirmed the idea of the organisation as it reaches 18\% to 25\% two seconds before the departure and 30\% to 50\% one second before the departure (Figure~\ref{fig:kendall_rank}). In the literature we found other cases of initiations: \cite{Leblond.2006, Pillot.2010} have shown that a three-spined stickleback \textit{Gasterosteus aculeatus} or a sheep \textit{Ovis aries} alone moving away from the herd can initiate a collective departure, \cite{Byrne.1990} have notice a large variety of intiations for groups of mountain baboons \textit{Papio ursinus} where the initiator can be joined by the group immediately or \cite{Leca.2003, Petit.2009} have observed for white-headed capuchins \textit{Cebus capucinus} a synchronization of their behaviours and a minimum proportion of the population is capable to launch a collective departure.\\

In conclusion, this study shows that the size of the population and the structure of the environment affect the motion of each individual in the groups and the group cohesion. The analysis of the dynamics shows that the zebrafish oscillate mainly in groups between the two patch in the environment and that the majority of the departures are collective. During the collective departures, we observe that an intra-group organisation appears prior to the transition. Increasing the population size makes this organisation less predictable. Finally, we noticed that a few seconds before the collective departures the groups has a particular spatial organisation.


\section{Methods}

\subsubsection{Fish and housing}

We bred 600 AB strain laboratory wild-type zebrafish (\textit{Danio rerio}) up to the adult stage and raised them under the same conditions in tanks of 3.5L by groups of 20 fish in a zebrafish aquatic housing system (ZebTEC rack from Tecniplast) that controls water quality and renew 10\% of water in the system every hour. Zebrafish descended from AB zebrafish from different research institutes in Paris (Institut Curie and Institut du Cerveau et de la Moelle \'Epini\`ere). AB zebrafish show zebra skin patterns and have short tail and fins. They measure about 3.5 cm long. All zebrafish used during the experiments were adults from 7 to 8 months of age. We kept fish under laboratory conditions: $27\,^{\circ}{\rm C}$, 500$\mu$S salinity with a 10:14 day:night light cycle, pH is maintained at 7.5 and nitrites (NO$^{2-}$) are below 0.3 mg/L. Zebrafish are fed two times a day (Special Diets Services SDS-400 Scientic Fish Food).

\begin{figure}[ht]
\centering
\includegraphics[width=0.5\textwidth]{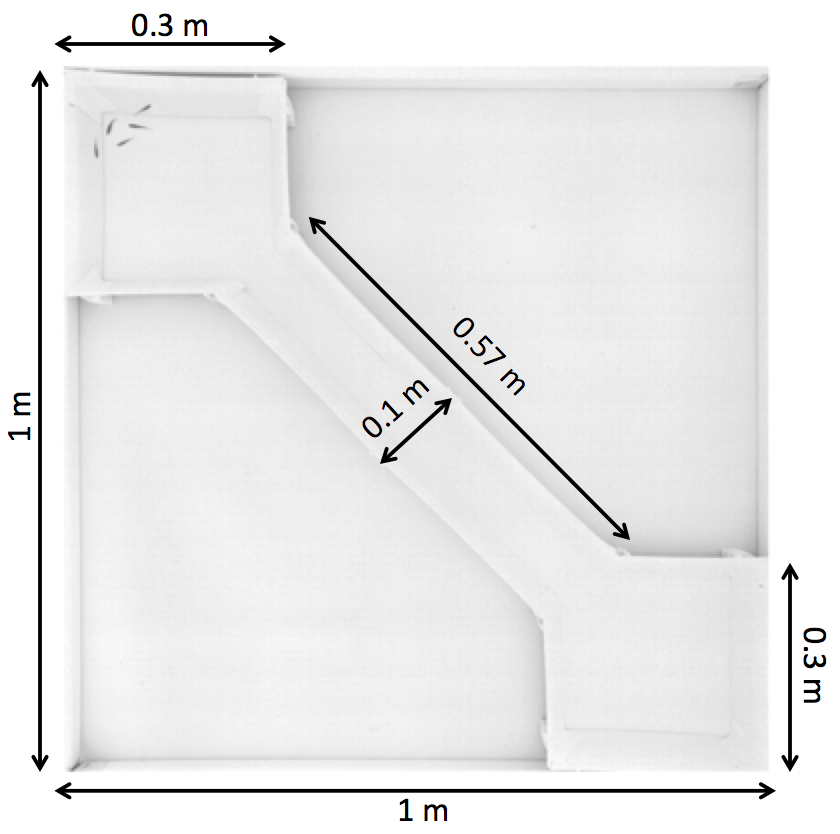}
\caption{\textbf{Experimental setup}.  A tank of 1 m x 1 m is divided into three areas: two rooms (0.3 m x 0.3 m) connected by a corridor (0.57 m x 0.1 m). The water column has a height of 6 cm. The luminosity is ensured by 4 LED lamps of $33 W$ (LP-500U) placed on  corners of the tank and directed towards the walls to provide indirect lighting. The whole setup is confined in a 2 m x 2 m x 2.35 m experimental chamber surrounded by white sheets to isolate the experiments and to homogenise luminosity.}
\label{fig:setup}
\end{figure}

\subsubsection{Experimental setup}

The experimental tank consists in a 1.2 m x 1.2 m tank confined in a 2 m x 2 m x 2.35 m experimental area surrounded by white sheets, in order to isolate the experiments and homogenise luminosity. A white opaque perspex frame (1 m x 1 m x 0.15 m - interior measures) is placed in the center of the tank. This frame help us to position the two rooms and the corridor. The squared rooms (0.3 m x 0.3 m) and the corridor (0.57 m x 0.1 m) have been designed on Computer-Aided Design (CAD) software and cut out from Poly(methyl methacrylate) (PMMA) plates of 0.003 m thickness. Each wall are titled, ($20^{\circ}$ from the vertical) to the outside with a vertical height of 0.14 m, to avoid the presence of blind zones for the camera placed at the vertical of the tank. The water column has a height of 6 cm, the water pH is maintained at 7.5 and Nitrites (NO$^{2-}$) are below 0.3 mg/L.
The experiments were recorded by a high resolution camera (2048 px x 2048 px, Basler Scout acA2040-25gm) placed above the experimental tank and recording at 15 fps (frame per second). The luminosity is ensured by 4 LED lamps of 33W (LED LP-500U, colour temperature: 5500 K - 6000 K) placed on each corner of the tank, above the aquarium and directed towards the walls to provide indirect lightning.

\subsubsection{Experimental procedure}

We recorded the behaviour of zebrafish swimming in the setup during one hour and did 12 replicates with groups of 1, 2, 3, 5, 7, 10 and 20 zebrafish. Every six replicates the setup was rotated by $90\,^{\circ}$ to prevent potential environmental bias (noise, light, vibrations, ...). Before each replicate, the starting chamber, from which the fish are released, is chosen randomly. Then, the fish are placed with a hand net in a cylindrical arena (20 cm diameter) made of Plexiglas in the centre the selected rooms. Following a five minutes acclimatisation period, this cylinder is removed and the fish are free to swim in the setup. The fish were randomly selected regarless of their sex and each fish was never tested twice to prevent any form of learning.

\subsubsection{Tracking \& data analysis}

Today, the studies on animal collective behaviours use methodologies based on massive data gathering, for exemple for flies (\textit{Drosophila melanogaster}) \cite{Branson.2009, Dankert.2009}, birds (\textit{Sturnus vulgaris}) \cite{Ballerini.2007, Miller.2007, Miller.2011}, fish (\textit{Notemigonus crysoleucas}) \cite{Strandburg-Peshkin.2013}.
Our experiments were tracked in real-time ("on-line") by a custom made tracking system based on blob detection. Each replicate except experiments with 20 zebrafish is also tracked  by post-processing ("off-line") with the idTracker software to identify each fish and their positions \cite{Perez.2014}. Each replicate consists of 54000 positions (for one zebrafish) to 1080000 positions (for 20 zebrafish). The idTracker software was not used for groups of 20 fish due to higher number of errors and too long computing time. For example, for a one hour video with 2 fish idTracker gives the results after 6 hours of processing and for a one hour video with 10 fish it lasts a week to do the tracking (with a Dell Percision T5600, Processor : Two Intel Xeon Processor E5-2630 (Six Core, 2.30GHz Turbo, 15MB, 7.2 GT/s), Memory : 32GB (4x8GB) 1600MHz DDR3 ECC RDIMM).

Since idTracker solved collisions with accuracy we calculated individual measures and characterised the aggregation level of the group. We also calculated the distances between each pair of zebrafish respectively, the travelled distances of each individual and their speeds. The calculation of the speed has been done with a step of a third of a second in sort of preventing the bias due to the tracking efficiency of idTracker that does not reach 100\% (see Table~\ref{fig:tracking_efficiency} of the appendix).

When all fish were present in the same room, we identified which fish initiates the exit from the room, established a ranking of exit for all the fish and calculated the distances between all zebrafish to the initiator to establish a ranking of distances. Finally, we confront these ranks and count the number of occurences for each ranking case. We check the results for different time steps before the initiation. The idea was to highlight a correlation between the spatial sorting and the ranks of exit and also a possible prediction of ranking of exit.

The Kendall rank correlation coefficient \cite{Kendall.1938}, $\tau$, is a measure of ordinal association between two measured quantities. It goes to 0 when the two quantities are independent and to 1 if they are correlated. It is compute by:

$\tau = \tfrac{\text{number} \text{ of} \text{ concordant} \text{ pairs } - \text{ number} \text{ of} \text{ discordant} \text{ pairs}}{\text{number} \text{ of} \text{ concordant} \text{ pairs } + \text{ number} \text{ of} \text{ discordant} \text{ pairs}}$.

Finally, we looked at majority events defined as the presence of more than 70\% of the zebrafish in one of the three areas of the setup, either in the room 1 or in the room 2 or in the corridor. To compute their numbers, we averaged the number of fish over the 15 frames of every second. This operation garanties that a majority event is ended by the departure of a fish and not by an error of detection during one frame by the tracking system. We then computed the durations of each of those events and counted the transitions from a room to the other one and sort them. All scripts were coded in Python using scientific and statistic libraries (numpy, pylab, scilab and matplotlib).

\subsubsection{Statistics}

For the Figure~\ref{fig:median_speed} we report the number of values of the speeds on the Table ~S\ref{fig:nb_values_speeds} of the appendix.
For the Figures~\ref{fig:median_speed}, ~\ref{fig:couples_boxplot} and ~\ref{fig:transitions} we tested the distributions using Kruskal-Wallis tests completed by a post-hoc test: Tukey's honest significant difference criterion.

\phantomsection

\section*{Animal ethics}

The experiments performed in this study were conducted under the authorisation of the Buffon Ethical Committee (registered to the French National Ethical Committee for Animal Experiments \#40) after submission to the state ethical board for animal experiments.

\section*{Data Availability}

Supporting data will be available on Dryad.

\section*{Competing interests}

We have no competing interests.

\section*{Authors' contributions}

AS carried out the lab work, the data analysis and the design of the set-up and drafted the manuscript; BC carried out the statistical analyses and participated in the data analyses and in the writing of the manuscript; LC developed the online tracker, stabilized the video aquisition and improved the idTracker; YC carried out the lab work, the design of the set-up and the data analysis; JH conceived the study, designed the study, coordinated the study and helped draft the manuscript. All authors gave final approval for publication.

\section*{Acknowledgments}

The authors thank Filippo Del Bene (Institut Curie, Paris, France) and Claire Wyart (Institut du Cerveau et de la Moelle \'Epini\`ere, Paris, France) who provided us the parents of the fish observed in the experiments reported in this paper.

\section*{Fundings}

This work was supported by European Union Information and Communication Technologies project ASSISIbf, FP7-ICT-FET-601074. The funders had no role in study design, data collection and analysis, decision to publish, or preparation of the manuscript.

\twocolumngrid

\clearpage
\onecolumngrid

\section{Supplementary information and figures}

Supplementary figures of "Loose social organisation of AB strain zebrafish groups in a two patches environment".

\subsection{From the individuals to the pair interactions}

\begin{table}[ht]
\centering
\begin{tabular}{|c|c|c|c|}
  \hline
  number of values & Room 1 & Room 2 & Corridor \\
  \hline
  1 fish & 54512 & 39829 & 24000 \\
  2 fish & 62941 & 61202 & 45698 \\
  3 fish & 129950 & 80216 & 73852 \\
  5 fish & 152880 & 119927 & 102647 \\
  7 fish & 253389 & 164149 & 170728 \\
  10 fish & 280719 & 218863 & 186999 \\
  \hline
\end{tabular}
\caption{\textbf{Number of values of speeds used for the Kruskal-Wallis and Tukey's honest significant difference criterion}. This table is related to Figure~\ref{fig:median_speed} of the article.}
\label{fig:nb_values_speeds}
\end{table}

\begin{figure}[ht]
\centering
\includegraphics[width=1\textwidth]{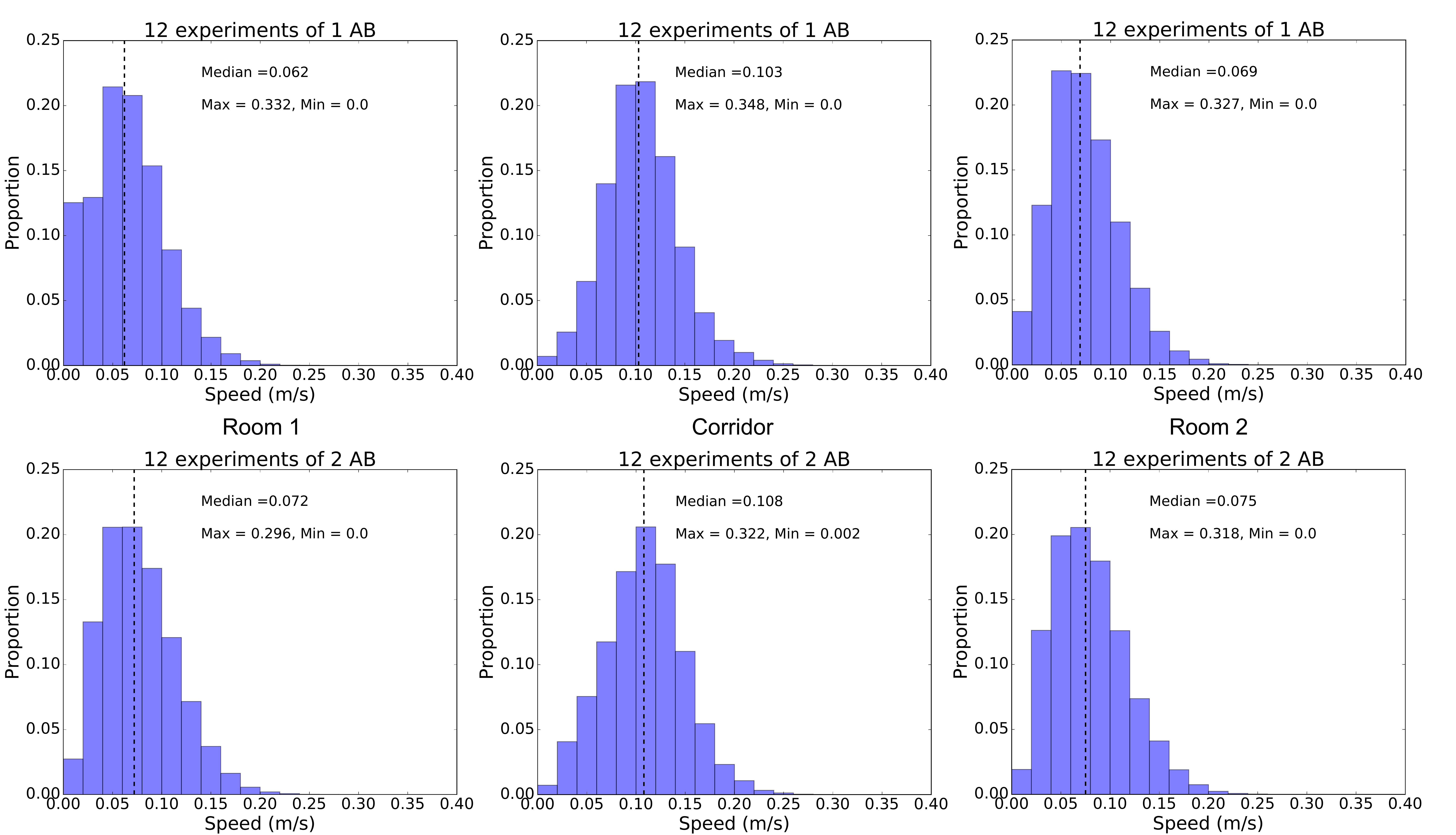}
\caption{\textbf{Distribution of the individual speeds for an individual and a pair of AB zebrafish.} }
 \label{fig:speed_1_2}  
\end{figure}

\begin{figure}[ht]
\centering
\includegraphics[width=1\textwidth]{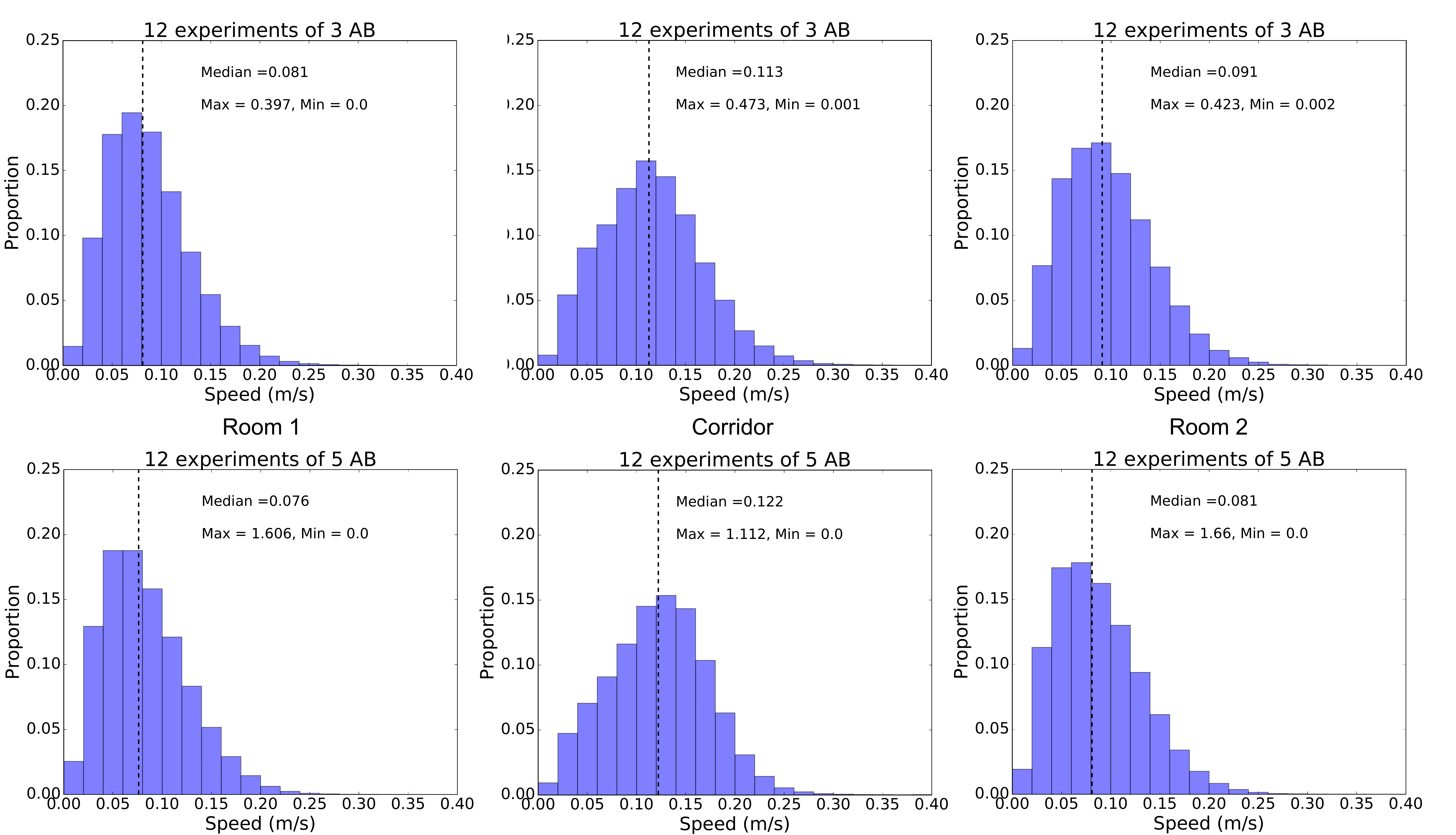}
\caption{\textbf{Distribution of the individual speeds for two group sizes.} Groups of 3 and 5 AB zebrafish.}
 \label{fig:speed_3_5}  
\end{figure}

\begin{figure}[ht]
\centering
\includegraphics[width=1\textwidth]{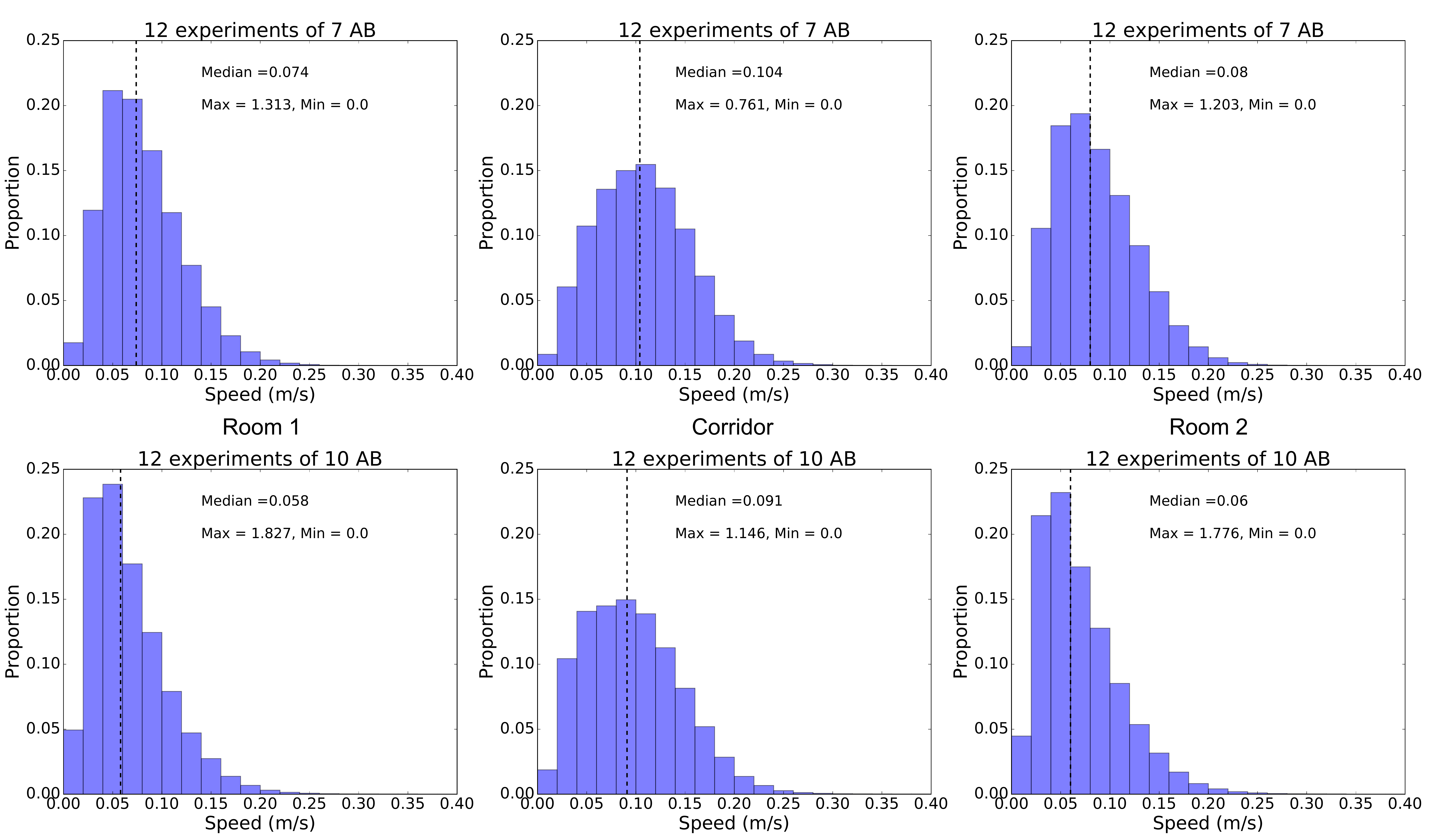}
\caption{\textbf{Distribution of the individual speeds for two group sizes.}  Groups of 7 and 10 AB zebrafish.}
 \label{fig:speed_7_10}  
\end{figure}

\begin{figure}[ht]
\centering
\includegraphics[width=0.80\textwidth]{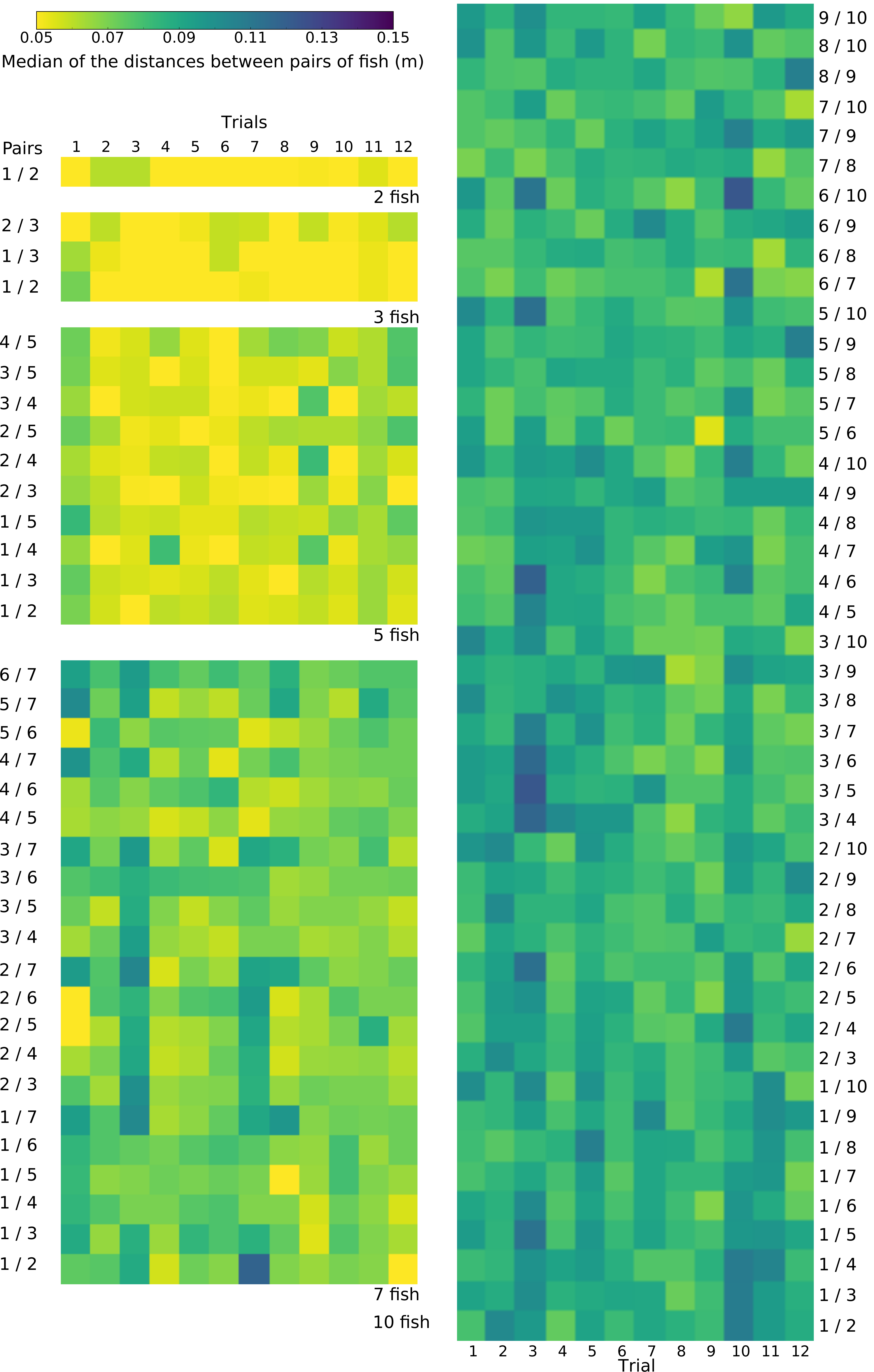}
 \caption{\textbf{Medians of the distances between all respective pairs of zebrafish per trial in the room 2 for 5 different group sizes.} }
 \label{fig:median_distances_pairs_2}
\end{figure}

\begin{figure}[ht]
\centering
\includegraphics[width=0.80\textwidth]{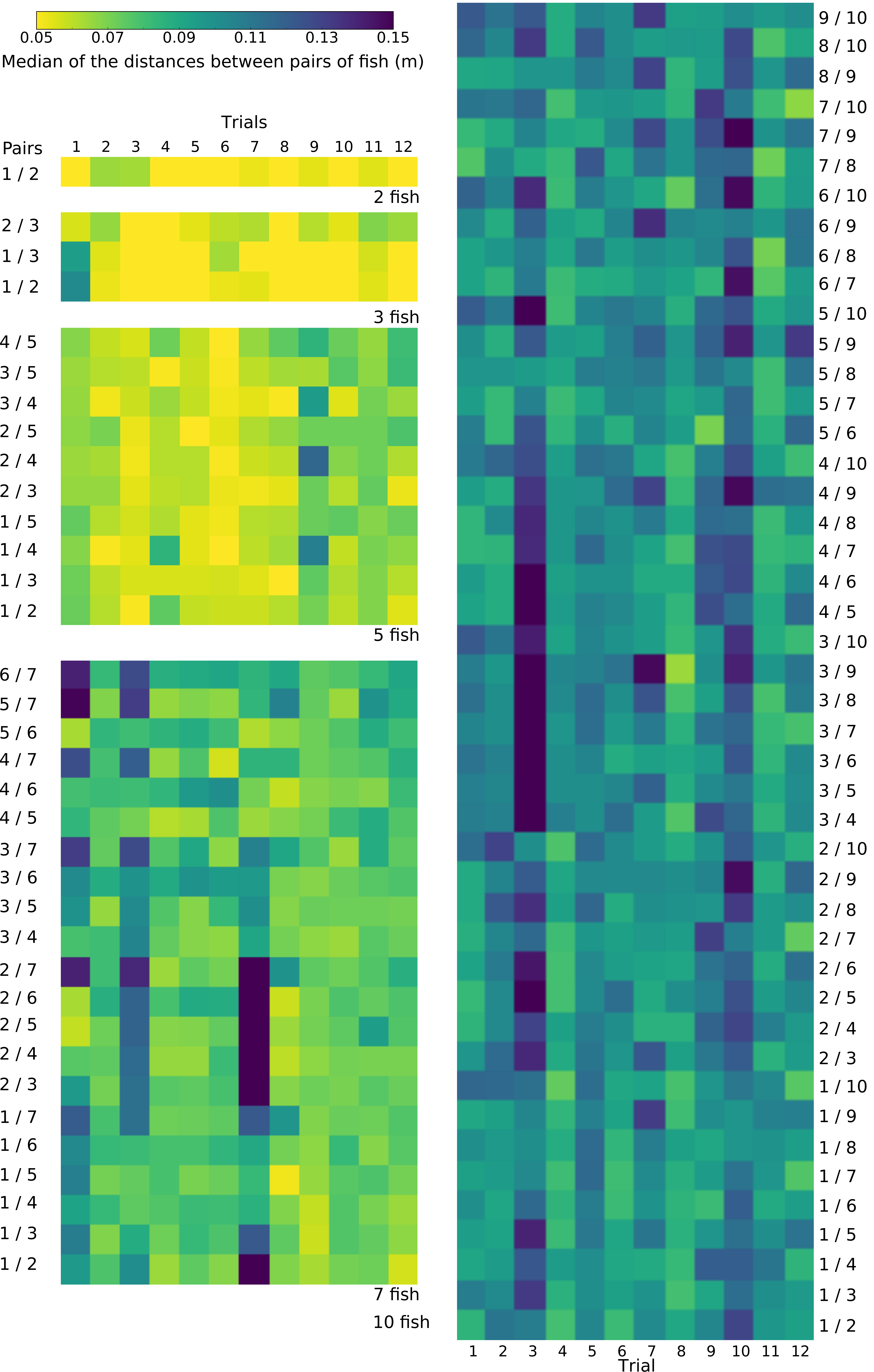}
 \caption{\textbf{Medians of the distances between all respective pairs of zebrafish per trial in the corridor for 5 different group sizes.} }
 \label{fig:median_distances_pairs_corridor}
\end{figure}

\begin{figure}[ht]
\centering
\includegraphics[width=.5\textwidth]{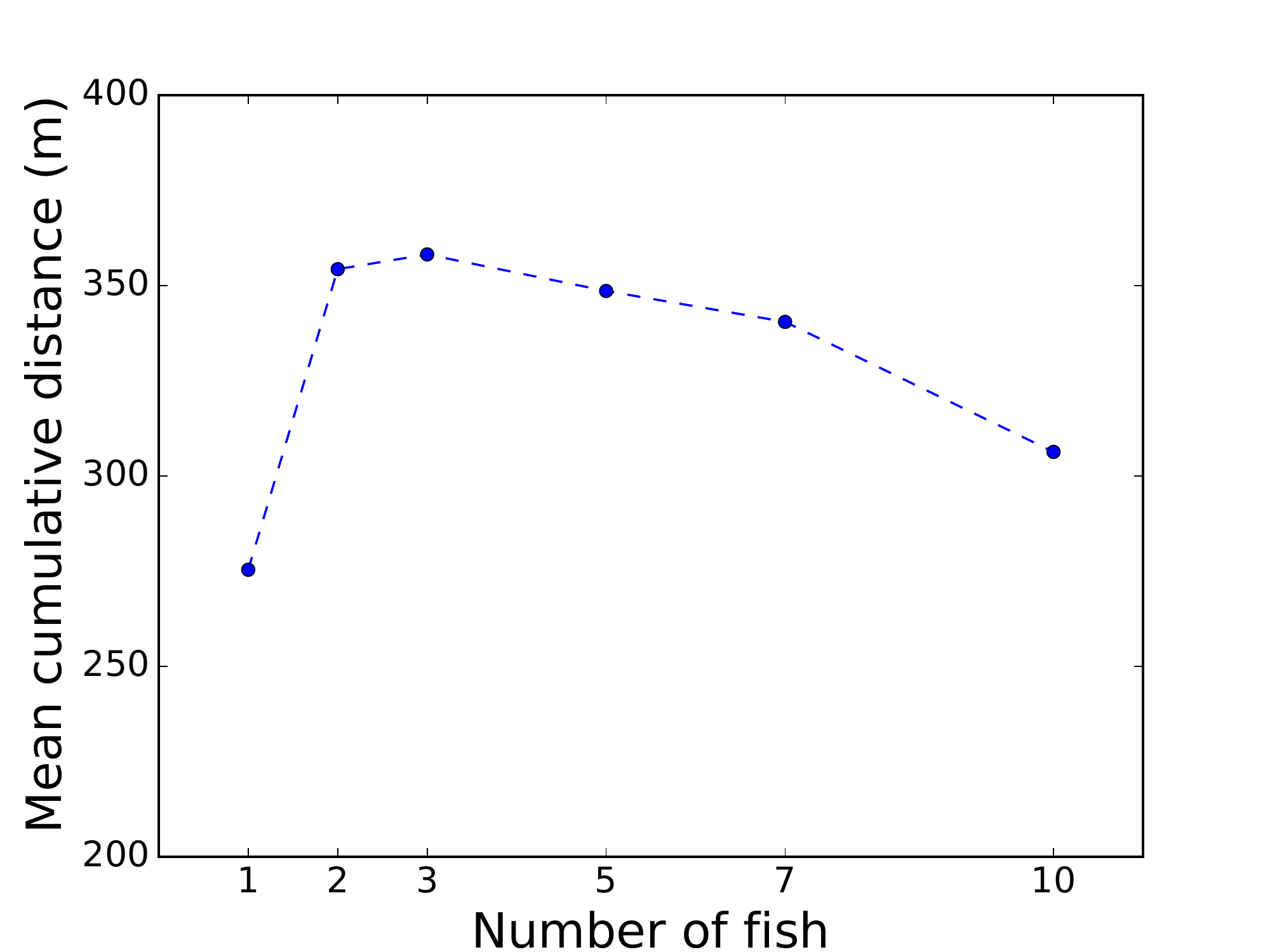}
\caption{\textbf{Median cumulative travelled distance for different group sizes.} The blue points represent the medians. Small groups of zebrafish travel more than bigger groups and fish alone. Kruskal-Wallis and Tukey's honest significant difference criterion show that the distributions of the travelled distances for fish alone and big groups (7 and 10 zebrafish) are not significantly different and likewise for groups of 2, 3, 5 and 7 zebrafish.}
 \label{fig:cumul_distance}  
\end{figure}

\clearpage
Figure~\ref{fig:cumul_distance} shows the median cumulative travelled distance by the zebrafish during one hour. It shows that a zebrafish alone travels the lowest distance, groups of 2 zebrafish travel the longest distance and then the bigger the group the shorter the distance travelled. We compared with a Kruskal-Wallis test the distributions of the travelled distances for all population sizes and found $p < 0.001$, which shows that at least one of the distribution is different  from the others. Table~\ref{fig:median_distance_compare} shows the results of the Tukey's honest significant difference criterion.

\begin{table}[ht]
\begin{tabular}{|c|c|c|c|c|c|c|c|}
  \hline
  & 1 fish & 2 fish & 3 fish & 5 fish & 7 fish & 10 fish \\
  \hline
 1 fish ‎ & - & $p < 0.001$ & $p < 0.001$ & $p < 0.001$ & ns & ns \\
  \hline
 2 fish ‎ & - & - & ns & ns & ns & $p < 0.001$ \\
  \hline
 3 fish ‎ & - & - & - & ns & ns & $p < 0.001$ \\
  \hline
 5 fish ‎ & - & - & - &- & ns & $p < 0.001$ \\
  \hline
 7 fish ‎ & - & - & - & - & - & $p < 0.001$ \\
  \hline
\end{tabular}
\caption{\textbf{Results of the Tukey's honest significant difference criterion between the medians of the travelled distances for every group sizes.} "ns" stands for non-significant.}
\label{fig:median_distance_compare}
\end{table}

\begin{figure}[ht]
\centering
\includegraphics[width=1\textwidth]{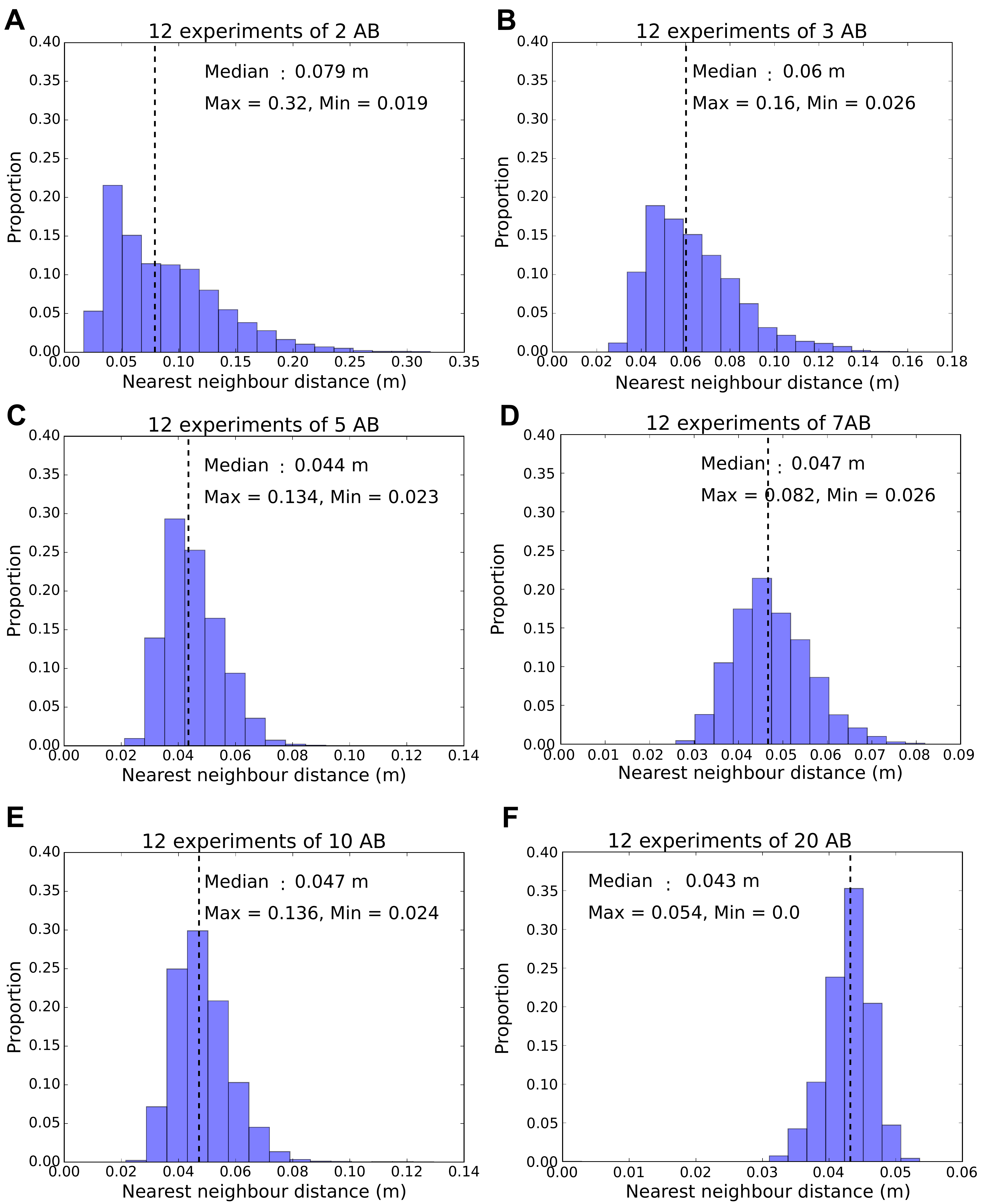}
  \caption{\textbf{Distributions of the nearest neighbour distances.} Groups of (A) 2 AB zebrafish, (B) 3 AB zebrafish, (C) 5 AB zebrafish, (D) 7 AB zebrafish, (E) 10 AB zebrafish and (F) 20 AB zebrafish. The plots are based on 648012 distances for 12 replicates. The dashed lines represent the medians. The distributions show for groups of 5, 7, 10 and 20 zebrafish similar medians and a shift to higher median values for smaller groups: 2 and 3 zebrafish. The nearest neighbour distances refer to the shortest distances between all zebrafish at every time step. It is a measure of group cohesion. The interest of such analysis is to dismiss the effect of the geometry of the set-up and to focus only on the group bearing.}
 \label{fig:all_neighbour}
\end{figure}

\clearpage

\subsection{Oscillations and collective departures}

\begin{figure}[ht]
\centering
\includegraphics[width=0.75\textwidth]{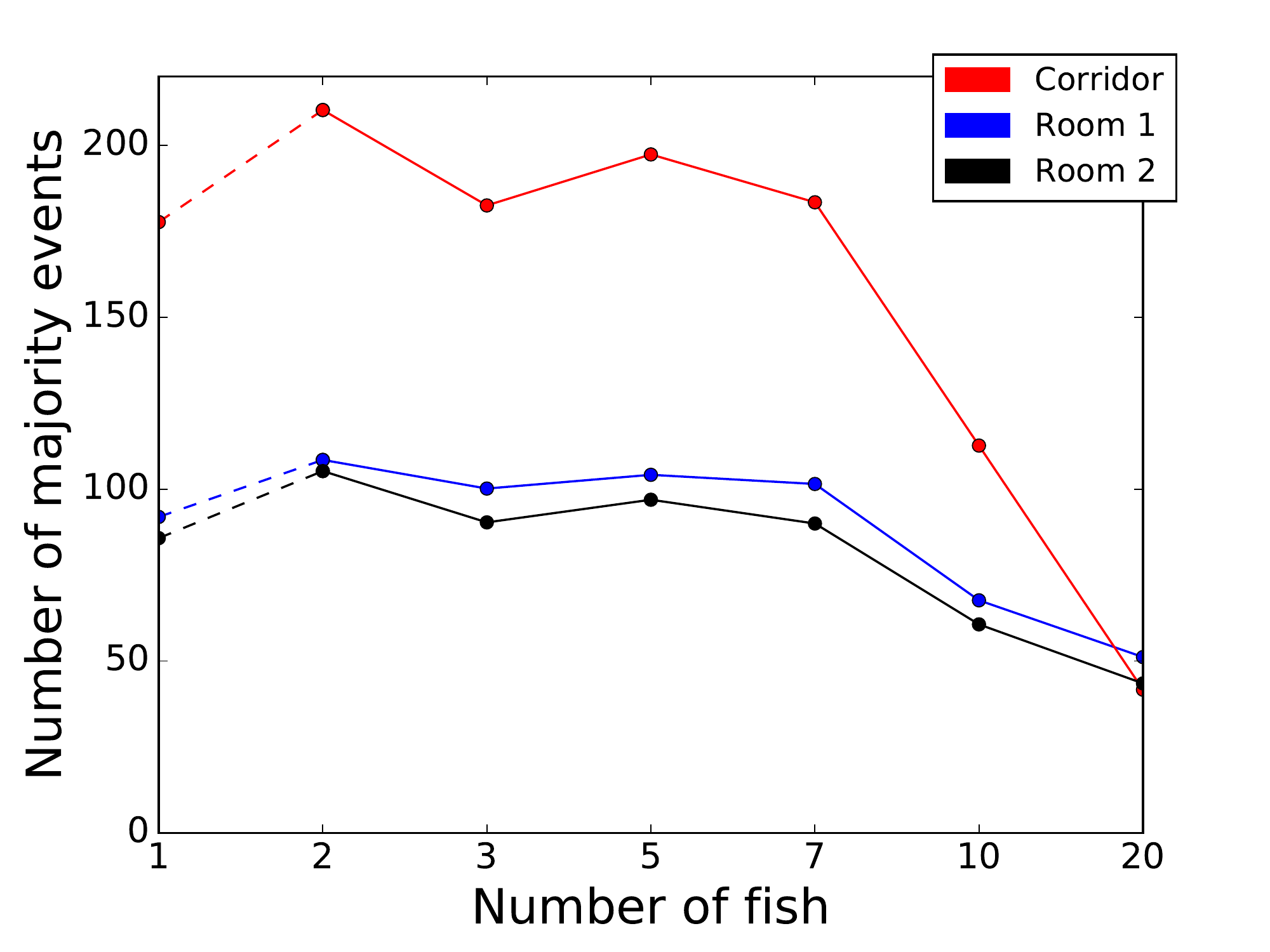}
 \caption{\textbf{Means of majority events with a majority of zebrafish in the three areas, for 7 group sizes} for 12 replicates each. To calculate the majority events, we count every time a majority of fish is located in one of the three areas. The red line indicates the number of majority events within the corridor, the blue line shows the number of majority events within the room 1 and the black line in the room 2. The dashed lines distinguish the experiments with one fish from the experiments with groups of fish. For all group sizes except for 20 zebrafish, there are more majority events in the corridor. The majority events are also very similar between room 1 and room 2. The number of majority events is relatively stable for groups of 1, 2, 3, 5 and 7 zebrafish in all three areas.}
 \label{fig:events}
\end{figure}

\begin{figure}[ht]
\centering
\includegraphics[width=0.75\textwidth]{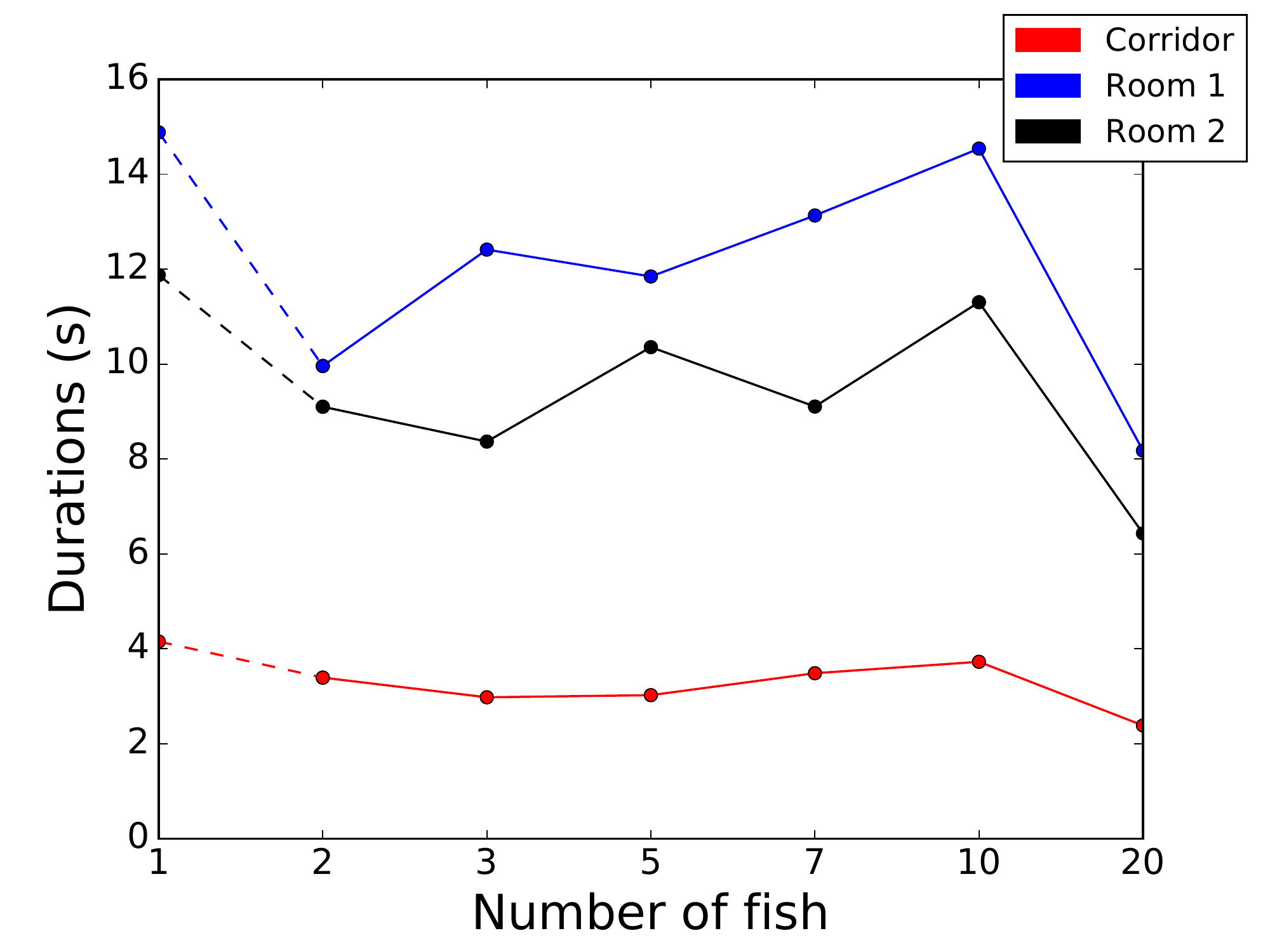}
 \caption{\textbf{Means of the time spent by a majority of fish in each area.} The results correspond to 12 replicates of 1 hour. The red line represents durations in the corridor, the blue line the durations in the room 1 and the black line the durations in the room 2. The dashed lines distinguish the experiments with one fish from the experiments with groups of fish. We show that the means of the durations are quite similar in rooms 1 and 2. They are shorter in the corridor than in both rooms. Increasing the size of the groups has no effect on the means of the durations in the corridor when it is generally followed by higher durations in both rooms. Finally, the durations strongly decrease when zebrafish are grouped by 20.}
 \label{fig:durations}
\end{figure}

To analyse the dynamics of the space occupancy in the set-up we computed the mean number of majority events and the mean durations and cumulative durations of occupancy by a majority of individuals within the three areas when a majority of the population is reached (Figure~S\ref{fig:events} and Figure~S\ref{fig:durations}). We define the majority as 70\% of the individuals being present in the considered section of the set-up. On the Figure~S\ref{fig:events}, we find more majority events in the corridor than in room 1 or room 2 except with groups of 20 zebrafish. Whatever the size of the group, we find almost the same number of majority events inside the rooms 1 and 2. Also, in all areas we see that for groups of 10 and 20 zebrafish the bigger the group the lower the number of majority events. The difference between the number of majority events in the corridor and in both rooms is relatively stable for groups of 1, 2, 3, 5 and 7 zebrafish but decreases when increasing the size of the groups (10 to 20 zebrafish). The mean number of majority events finally reaches almost the same value when 20 zebrafish are tested in the setup (room1: 51,2; room 2: 43.5; corridor: 41.7). Table~S\ref{fig:std_events} of the appendix (A) shows, for the 12 replicates of each group size, the standard deviations linked with the number of majority events (related to Figure~S\ref{fig:events}). On the Figure~S\ref{fig:durations}, we see that the means of the durations of the majority in each area follow a similar trend in both rooms and are longer than in the corridor. Increasing the size of the group has almost no effect on the durations in the corridor when it has an impact in the rooms, where fish stay longer in majority if the population size increases. However, for 20 zebrafish, durations decrease in all areas. Table~S\ref{fig:std_events} (B) of the appendix shows the standard deviations related to the Figure~S\ref{fig:durations}.

\begin{figure}[ht]
\centering
\includegraphics[width=1\textwidth]{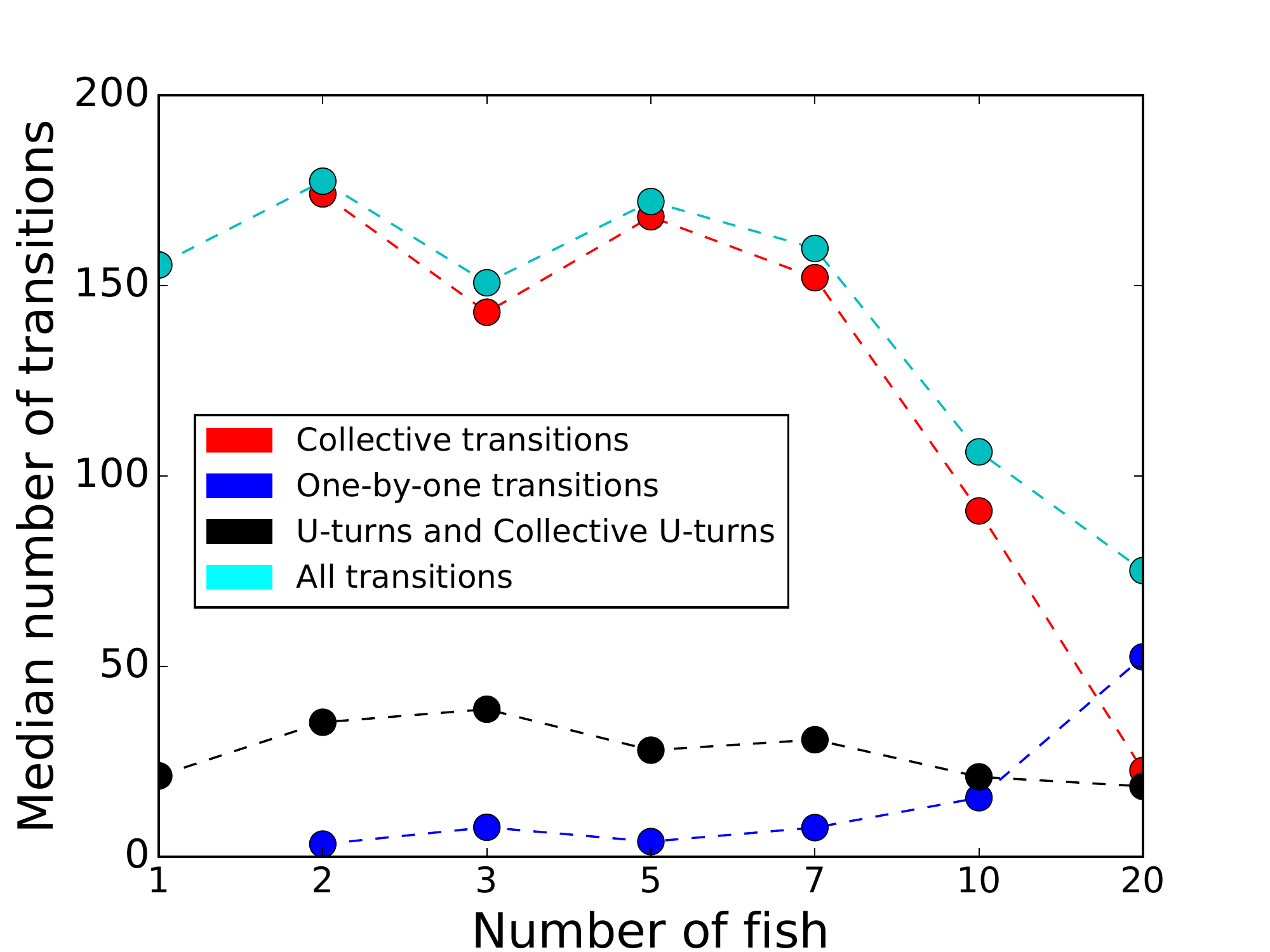}
 \caption{\textbf{Mean number of transitions for different group sizes.} The red curve shows "Collective transitions", the blue curve shows "One-by-one transitions", the black curve represents the "Collective U-turns" and the magenta ("All transitions") is the sum of "Collective transitions" and "One-by-one transitions". "One-by-one transitions" occur when the fish transit one by one from one room to the other. "Collective transitions" appear when the group transit between both rooms through the corridor. "Collective U-turns" occur when the group go back to the previous room. The figure shows that increasing the group sizes makes the number of "Collective U-turns" and "Collective transitions" decrease and the number of "One-by-one transitions" increase.}
 \label{fig:mean_transitions}  
\end{figure}

\begin{table}[ht]
\begin{tabular}{|c|c|c|c|c|c|c|c|}
  \hline
Group size & 1 fish & 2 fish & 3 fish & 5 fish & 7 fish & 10 fish & 20 fish \\
  \hline
Mean tracking & * & *& *& *& *& *& \\ efficiency & 100\% & 97\% & 98\% & 89\% & 94\% & 75\% & 94\% \\
  \hline
\end{tabular}
\caption{\textbf{Mean tracking efficiency.} The experiments are tracked by the idTracker program to find the individual identities and the positions of the fish \cite{Perez.2014}.}
\label{fig:tracking_efficiency}
\end{table}

\begin{table}[ht]
\centering
\begin{tabular}{|c|c|c|c|c|c|c|c|}
  \hline
  (A) Majority events & 1 fish & 2 fish & 3 fish & 5 fish & 7 fish & 10 fish & 20 fish \\
  \hline
  Std. corridor & 62.5 & 42.4 & 35.7 & 41.3 & 33.3 & 49.4 & 19.2 \\
  Std. room 1 & 30.2 & 17.4 & 18.8 & 21.0 & 14.5 & 26.1 & 14.3 \\
  Std. room 2 & 33.0 & 28.0 & 19.8 & 21.3 & 19.8 & 24.6 & 15.8 \\
  \hline
\end{tabular}
\begin{tabular}{|c|c|c|c|c|c|c|c|}
  \hline
  (B) Durations (s) & 1 fish & 2 fish & 3 fish & 5 fish & 7 fish & 10 fish & 20 fish \\
  \hline
  Std. corridor & 1.73 & 1.94 & 2.09 & 1.76 & 2.48 & 2.92 & 1.92 \\
  Std. room 1 & 21.7 & 10.2 & 11.8 & 10.9 & 12.2 & 13.2 & 8.28 \\
  Std. room 2 & 13.2 & 7.03 & 8.73 & 8.63 & 8.68 & 10.7 & 6.57 \\
  \hline
\end{tabular}

\begin{tabular}{|c|c|c|c|c|c|c|c|}
  \hline
  (C) Transition types & 1 fish & 2 fish & 3 fish & 5 fish & 7 fish & 10 fish & 20 fish \\
  \hline
  Collective & 32.9 & 23.1 & 16.4 & 19.5 & 14.0 & 22.4 & 6.9 \\
  One-by-one & 0.1 & 2.4 & 3.7 & 3.3 & 3.7 & 7.8 & 8.8 \\
  U-turns & 12.0 & 12.3 & 7.0 & 6.9 & 5.3 & 5.4 & 4.5 \\
  \hline
\end{tabular}
\caption{\textbf{Standard deviations of the means of (A) majority events, (B) durations, (C) numbers of transition types with a majority of zebrafish in the three sections of the setup}.}
\label{fig:std_events}
\end{table}

\begin{figure*}[ht]
\centering
\includegraphics[width=1\textwidth]{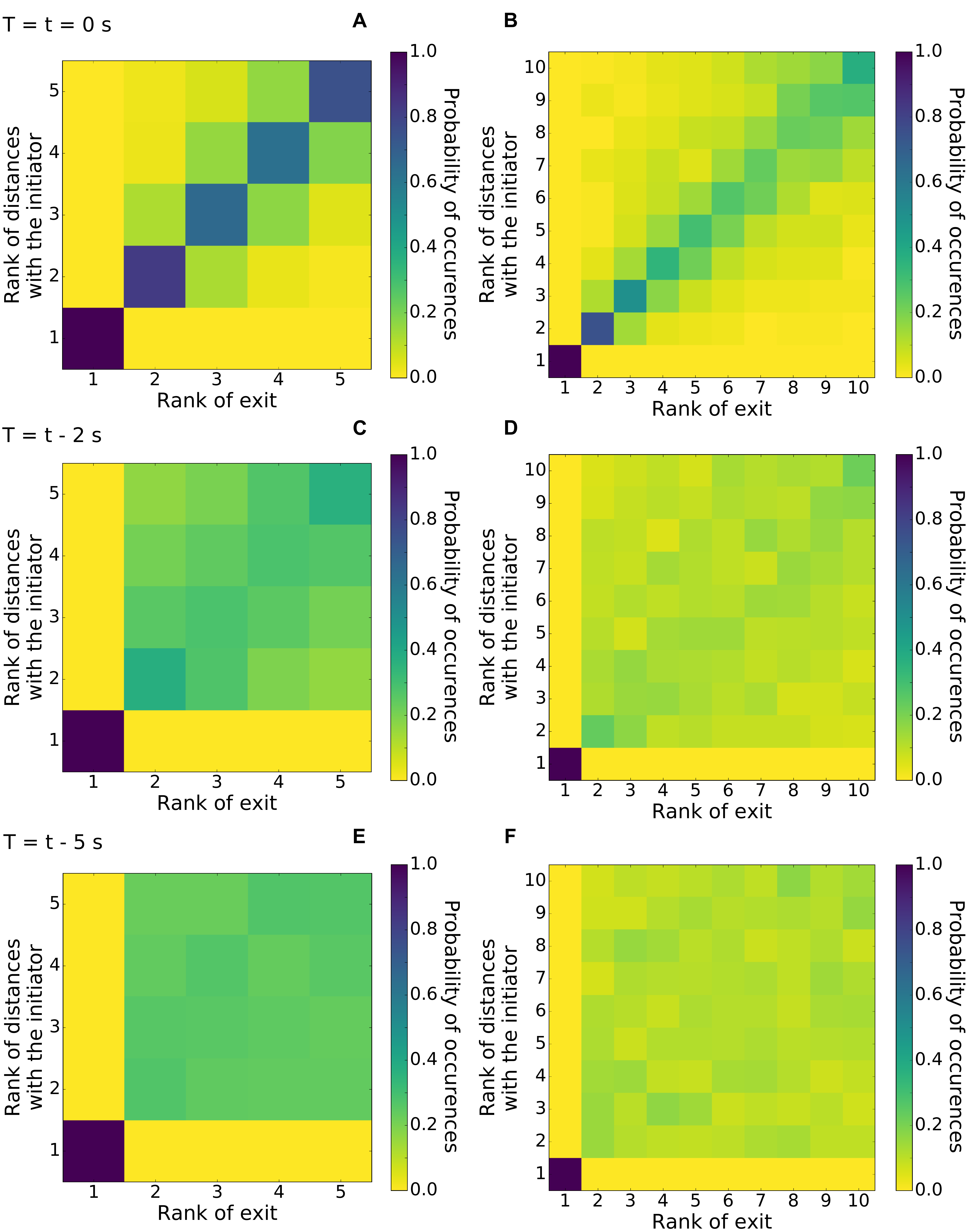}
 \caption{\textbf{Probability of occurrence of the rank of exit with the rank of distances from the initiator.} The results correspond to groups of 5 zebrafish (left column) and 10 zebrafish (right column). We counted N = 1456 exits for 12 replicates with 5 zebrafish and N = 277 for 12 replicates with 10 zebrafish. (A) and (B) show the map at the time where the initiator leave the room, (C) and (D) 2 seconds before, (E) and (F) 5 seconds before. As an example, in (A) the probability of occurrence where the second fish leaves the room and has the shortest distance from the initiator is 0.82. As an example, (A) fish with rank of 2 for exit and for distances (closest distance with the initiator) show a probability of 0.82 to be the closest fish to the initiator. This probability decreases to 0.12 for fish with rank of 2 for exit and rank of 3 for distances (the second closest distance with the initiator). Focusing now on (C), 2 seconds before the initiation: the first probability decreases from 0.82 to 0.37 when the second one increases from 0.12 to 0.26. Plots for experiments with 3 and 7 zebrafish are also in the annexe Figure~S\ref{fig:map_sortie_3_7}.}
 \label{fig:map_sortie_5_10}
\end{figure*}

\begin{figure}[ht]
\centering
\includegraphics[width=0.9\textwidth]{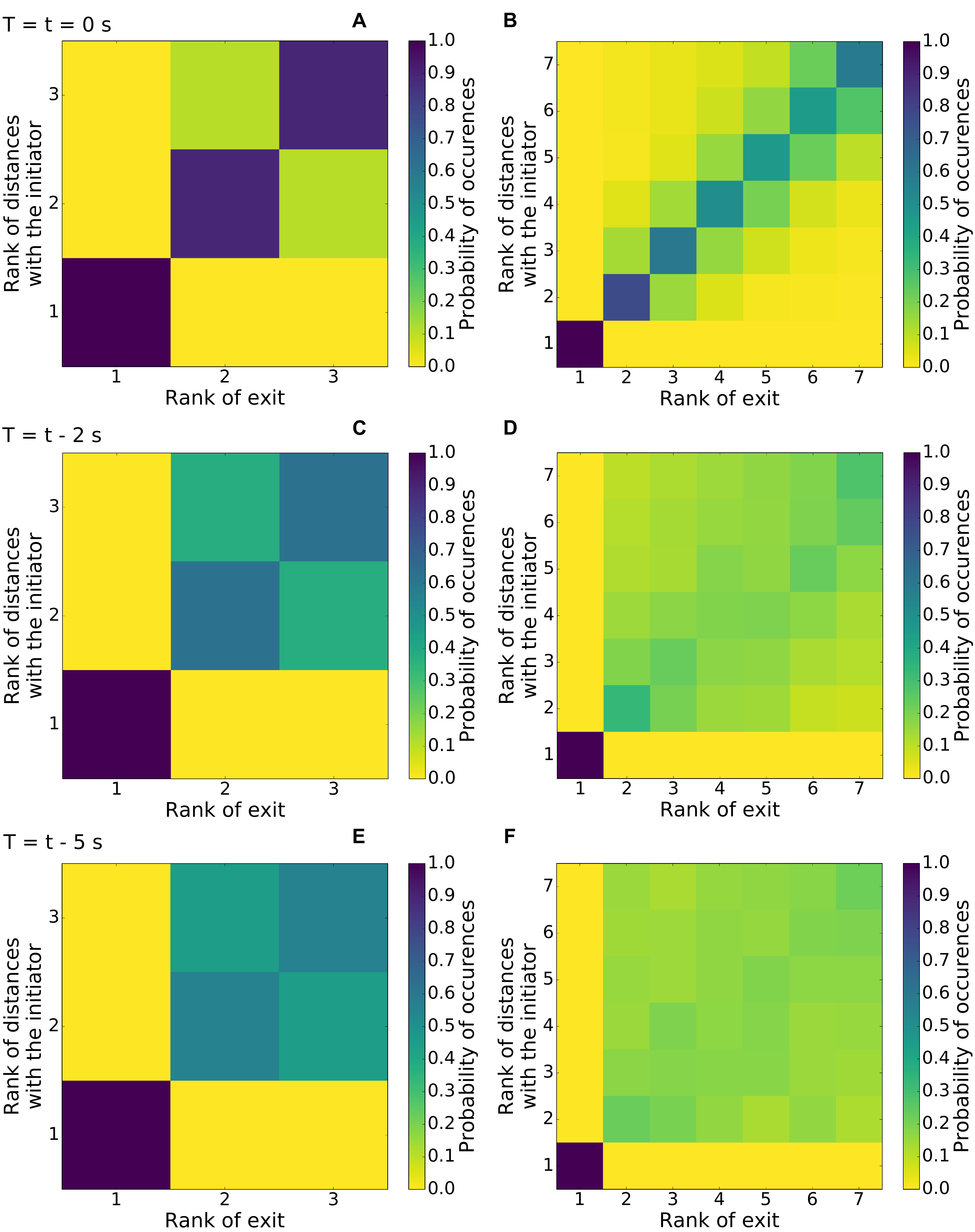}
 \caption{\textbf{Probability of occurrence of the rank of exit with the rank of distances from the initiator.} The results correspond to groups of 3 zebrafish (left column) and 7 zebrafish (right column). We counted N = 2195 exits for 12 replicates with 3 zebrafish and N = 1020 for 12 replicates with 7 zebrafish. (A) and (B) show the map at the time where the initiator leave the room, (C) and (D) 2 seconds before, (E) and (F) 5 seconds before.}
 \label{fig:map_sortie_3_7}
\end{figure}


\end{document}